\documentclass[final,5p,times,twocolumn]{elsarticle}

\usepackage{graphicx}
\graphicspath{{./figs/}}
\usepackage{amsmath}
\usepackage{amssymb}
\usepackage{subfig}
\usepackage{stfloats}
\usepackage[utf8]{inputenc}

\usepackage{mathtools, cuted}
 
\newcounter{bla}

\journal{Computer Physics Communications}

\usepackage{hyperref}
\hypersetup{
    colorlinks=true,
    linkcolor=blue,
    filecolor=blue,      
    urlcolor=blue,
}

\begin{document}

\begin{frontmatter}

\title{Merlin++, a flexible and feature-rich accelerator physics and particle tracking library}
\author[a,b]{Robert~B.~Appleby}
\author[c]{Roger~J.~Barlow\corref{author}}
\author[d]{Dirk~Kr\" ucker}
\author[e]{James~Molson}
\author[c]{Scott~Rowan}
\author[a]{Sam~Tygier}
\author[e]{Haroon~Rafique}
\author[d]{Nicholas~Walker}
\author[f,b]{Andrzej~Wolski}
 
\cortext[author]{\hspace{1pt} Corresponding author.\\\indent\hspace{10pt}\textit{E-mail address:} \href{mailto:roger.barlow@cern.ch}{roger.barlow@cern.ch} (R. J. Barlow)}
\address[a]{The University of Manchester, Oxford Rd, Manchester M13 9PL, UK}
\address[b]{The Cockcroft Institute, Keckwick Ln, Daresbury, Warrington, WA4 4AD, UK}
\address[c]{The University of Huddersfield, Queensgate, Huddersfield, HD1 3DH, UK}
\address[d]{DESY, Notkestra\ss e 85, D22607 Hamburg, Germany}
\address[e]{CERN, 
CH1211 Gen\`eve 23, Switzerland}
\address[f]{The University of Liverpool, Liverpool L69 7ZX, UK}

\begin{abstract}
Merlin++ is a C++ charged-particle tracking library developed for the simulation and analysis of complex beam dynamics within high energy particle accelerators. Accurate simulation and analysis of particle dynamics is an essential part of the design of new particle accelerators, and for the optimization of existing ones. Merlin++ is a feature-full library with focus on long-term tracking studies. A user may simulate distributions of protons or electrons in either single particle or sliced macro-particle bunches. The tracking code includes both straight and curvilinear coordinate systems allowing for the simulation of either linear or circular accelerator lattice designs, and uses a fast and accurate explicit symplectic integrator. Physics processes for common design studies have been implemented, including RF cavity acceleration, synchrotron radiation damping, on-line physical aperture checks and collimation, proton scattering, wakefield simulation, and spin-tracking. Merlin++ was written using C++ object orientated design practices and has been optimized for speed
using multicore processors. This article presents an account of the program, including its functionality and guidance for use.

{\bf Program Summary}

{\em Program Title}: Merlin++

{\em Program Files DOI}: \href{https://dx.doi.org/10.5281/zenodo.3700155}{10.5281/zenodo.3700155}

{\em Licensing provisions}: GPLv2+ 

{\em Programming language}: C++        

{\em Nature of problem}: Complexity of particle accelerators beam dynamics over extensive tracking distances.

{\em Solution method}: Long-term particle accelerator and tracking simulations utilizing explicit symplectic integrators.

{\em Additional comments}: For further information see \href{https://github.com/Merlin-Collaboration}{github.com/Merlin-Collaboration}
\end{abstract}


\end{frontmatter}


\section{Introduction}

Modern high energy particle accelerators are large and complex, and beam dynamics simulations are therefore fundamental to their design and optimization. 

A charged particle accelerator is designed as a lattice structure of electromagnetic accelerating and/or focusing components. 
An initial design may consist of standard arrangements of common elements such as transverse-magnetic (TM) mode radio-frequency (RF) resonating cavities to transfer energy and accelerate particles, and dipole and quadrupole electromagnets to guide and focus the beam. A more refined design is then achieved through an iterative process of beam dynamics simulations wherein lattice element parameters are altered and higher-order magnetic components such as sextupoles and octupoles are included where necessary to account for non-linear instabilities. 
Specific lattice sections may be designed for specific purposes: for example, a lattice subsection of bending dipoles may be constructed symmetrically in a double achromat so that there is zero energy dispersion in the beam at either end, or a triplet of strong quadrupole magnets may be used to focus the beam at a collision point.

Successful operation of an accelerator relies on beam stability and field synchronisation. For example, in linear accelerators, with aligned RF cavities in series, synchronization of AC voltage signals with incident particles is vital in achieving acceleration. 
For circular accelerators, avoiding disruptive resonances is 
important to ensure long-term stability, as losses may only develop over thousands or even millions of turns.
Due to the scale and complexity of modern accelerator lattices, accurate beam dynamics simulations are crucial in optimizing the dynamic aperture and beam lifetime. 

Real world lattice elements will not match the optimal design exactly, being subject to alignment and field strength errors. 
Simulations are also necessary to ensure that any loss of performance due to such variation falls within allowable limits. 
A simulation is then performed not just once but many times, so the program needs to be optimized for speed.

Merlin++, described in this article, is such a simulation program. 
After a summary of existing tracking codes
and how Merlin++ compares with them in
Section~\ref{sec:codes}
the physical processes modelled are described in Section~\ref{sec:overview}
and
software issues in Section~\ref{sec:software}.
Section~\ref{sec:accuracy} contains 
results from benchmarking Merlin++ against data and  other codes, 
and Section~\ref{sec:speed} outlines the
measures used to optimize the speed of the program. 
Section~\ref{sec:using} shows a potential user how to get started, and finally some
suggestions for possible future development are given in 
Section~\ref{sec:conclusions}.

This article is intended primarily for readers who
need to simulate an accelerator and are considering
whether Merlin++ will provide what they require: a knowledge of basic accelerator concepts is therefore assumed. Any other readers (perhaps software specialists) can, if interested, find explanations in any basic accelerator textbook, such as~\cite{basic}.

\section{Tracking Codes and Merlin++}

\label{sec:codes}
In recent decades, significant advances have been made in the accuracy and stability of mathematical methods used by particle tracking codes. Historically, so-called matrix codes such as TURTLE\cite{turtle}, which used Brown and Rothacker's Transport \cite{Transport}, used the thin-lens approximation with transfer matrices propagating a particle through an accelerator lattice along a path close to a reference orbit.
Such codes are inherently limited due to truncation of the maps to a specific order of non-linearity, typically 3rd-order. More accurate codes, such as Cosy-Infinity \cite{cosyinfinity} which uses differential algebra to generate maps to arbitrary order, can be used for tracking with a distribution of particles with varying particle energies/orbits, however, this method very quickly becomes impractical for large lattices and long simulations due to computational limitations.  Ruth \cite{ruth} proposed a more practical approach, involving the derivation of explicit equations for each component from integrable component-specific Hamiltonians. This allows for accurate stepwise propagation through each element. Such  integrators are possible due to most common accelerator component Hamiltonians being integrable, either directly or when split. More advanced Hamiltonian splitting methods for various components were further developed by Forest and Ruth \cite{ruthforest}, Yoshida \cite{yoshida}, Wu, Forest and Robin\cite{WuForestRobin} and Laskar and Robutel \cite{LaskarRobutel}.

Various beam dynamics and accelerator physics codes have been developed over the years to assist in the accelerator design process. 
 For example MAD \cite{MAD}, is widely used for single particle dynamics simulations, Bmad \cite{Bmad}, to study non-linearities and collective effects, FLUKA \cite{FLUKA2} for detailed particle-matter interactions, and there are many more. Historically, whenever the scope of a research study exceeded that of existing codes, designers had to create their own code base. The programming languages and practices utilized have not generally been sustainable, so many tracking codes have been developed, used and abandoned. Due to lack of an alternative some code bases are maintained even with known core design issues and/or being built in an outdated or inefficient language. Moreover a new accelerator simulation will require consideration not only of magnets and RF cavities but of other physical processes which influence particles, such as synchrotron radiation and collimation.   A typical existing tracking code is limited in the additional processes that may be included in the simulation, and in many cases, due to being later designed as an addition/attachment to a legacy code base, in these codes such features have computationally expensive implementations.

Merlin++ was therefore architecturally designed as expandable and general-purpose. This was achieved using the C++ language. C++ is a high-level object-orientated language which crucially does not compromise low-level access and optimization features. Utilizing the full feature set of the language, from C++11 and beyond, such as generic programming, polymorphism and memory allocation, as well as cutting-edge technologies such as move semantics and native multi-threading, Merlin++ is designed from the ground up to be accessible, modular and fast. 

\begin{figure}[h]
  \includegraphics[width=0.48\textwidth]{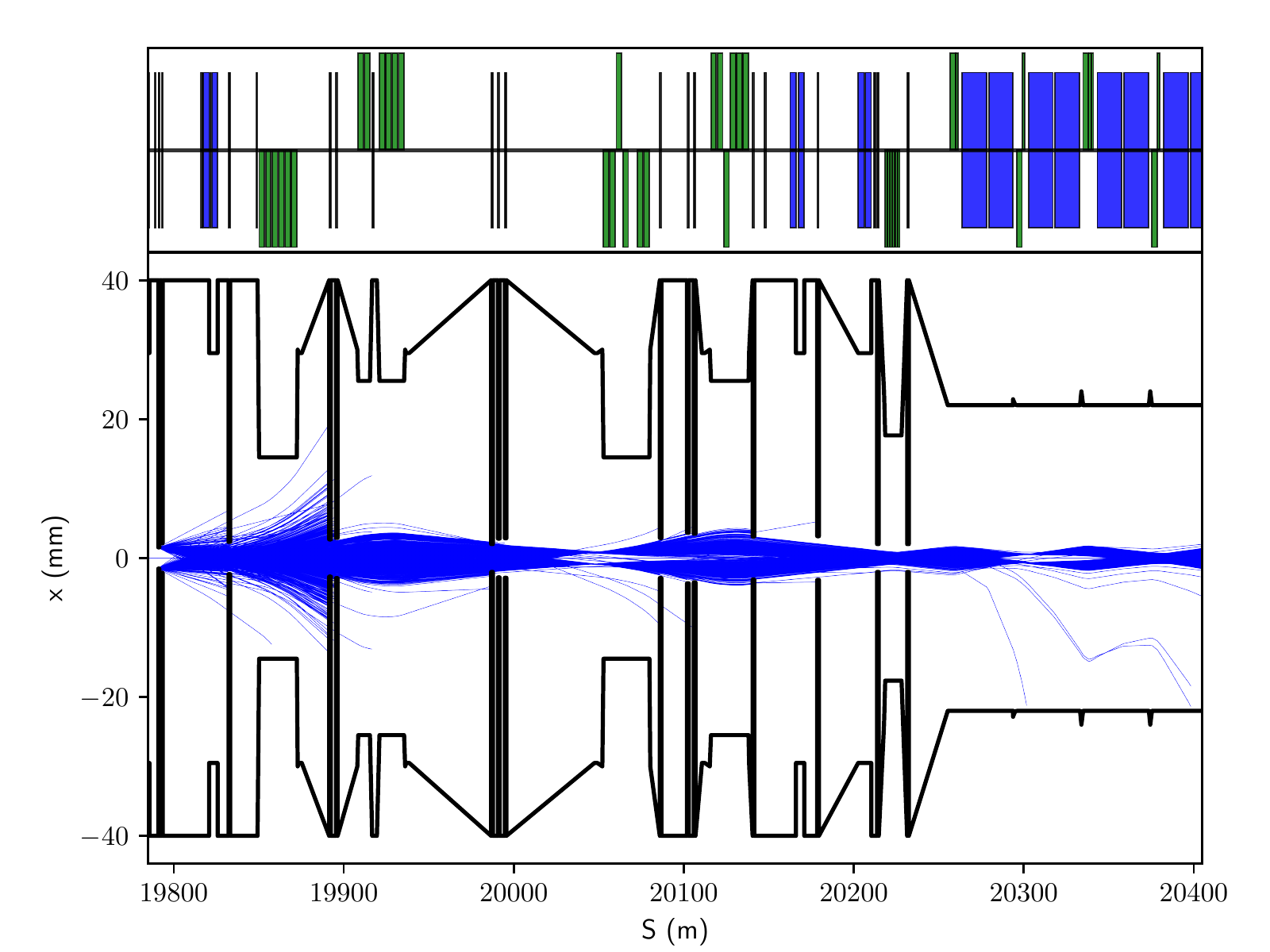}
  \caption{Scattered particles in the collimation region of the LHC. Upper plot shows the arrangement of dipoles (blue), quadrupoles (green) and collimators (black). Lower plot shows the machine aperture (black) and proton tracks (blue).}
\label{fig:IR7}
\end{figure}

Merlin++ has been used and developed for some years and many additional features and functionality have easily been incorporated by new user-developers without affecting the existing code. It has grown to be a powerful and  feature-rich general purpose accelerator physics library. An example of a process added to Merlin++ in recent years is the collimation and scattering feature set, 
including several different scattering process such as multiple-Coulomb and Rutherford scattering as well as  models for elastic proton-nucleus and single diffractive proton-nucleon interactions. 
Figure \ref{fig:IR7} shows an example: the Merlin++ output of particles interacting in the collimation region of the LHC. It shows how several halo particles have undergone scattering in the collimator material, but have not been absorbed and continued to propagate and subsequently impacted other components of the machine. This type of simulation is important for machine protection studies. These new physics processes were readily incorporated into the existing framework. 

\section{Overview of Merlin++}

\label{sec:overview}
Merlin++~\cite{merlin_zenodo} is a 
C++ physics library 
for high energy particle accelerator systems design and particle tracking simulations. Its functionality is similar to that of MAD, but with many more features, and is comparable in many ways to SixTrack~\cite{sixtrack}. Over the years, Merlin++ has been extended and applied to a variety of use-cases, such as ILC beamline and damping rings studies \cite{merlin_ilcdfs,merlin_ilcground,merlin_ilcpackage}, NLC depolarisation/spin-tracking \cite{merlin_nlcspin}, and LHC collimation \cite{merlin_collimation2010,merlin_collimation2012,merlin_collimation2014, merlin_collimation2015,merlin_collimation2016,merlin_collimation2017}. Currently, Merlin++ remains under active development and use for HL-LHC, FCC~\cite{rafique_proton_2017} and SppC collimation studies~\cite{yang_collimation_2019}. As a result, the code base has grown to 
become one of the most feature-rich and functionally capable tracking codes available.

It provides a 6-dimensional phase-space tracking library with multiple tracking integrators, including symplectic and 1\textsuperscript{st} and 2\textsuperscript{nd}-order transport integrators. 
Bunches of 
protons or electrons can be tracked along either linear or circular accelerator lattice structures. Particle bunches may be comprised of a distribution of single particles or a collection of sliced macro-particles. 
Optional physics processes are also provided, including RF cavity acceleration, synchrotron radiation, on-line physical aperture checks, collimation, proton diffractive scattering, wake-field simulation and spin-tracking. 

\subsection{Architecture and Design Philosophy}

Merlin++'s structure was developed with two fundamental design philosophies:
\begin{enumerate}
    \item To be functional and of high performance while being easy to learn and flexible.
    \item To be comprised of loosely-coupled and independently maintainable modules.
\end{enumerate}

\noindent Accordingly 
it consists of four loosely-coupled and extensible modules:
\begin{enumerate}
    \item Lattice Structure
    \item Beam Parameters
    \item Particle Tracking
    \item Physics Processes
\end{enumerate}

\begin{figure*}[t]
\centering
\includegraphics[width=1\textwidth,clip]{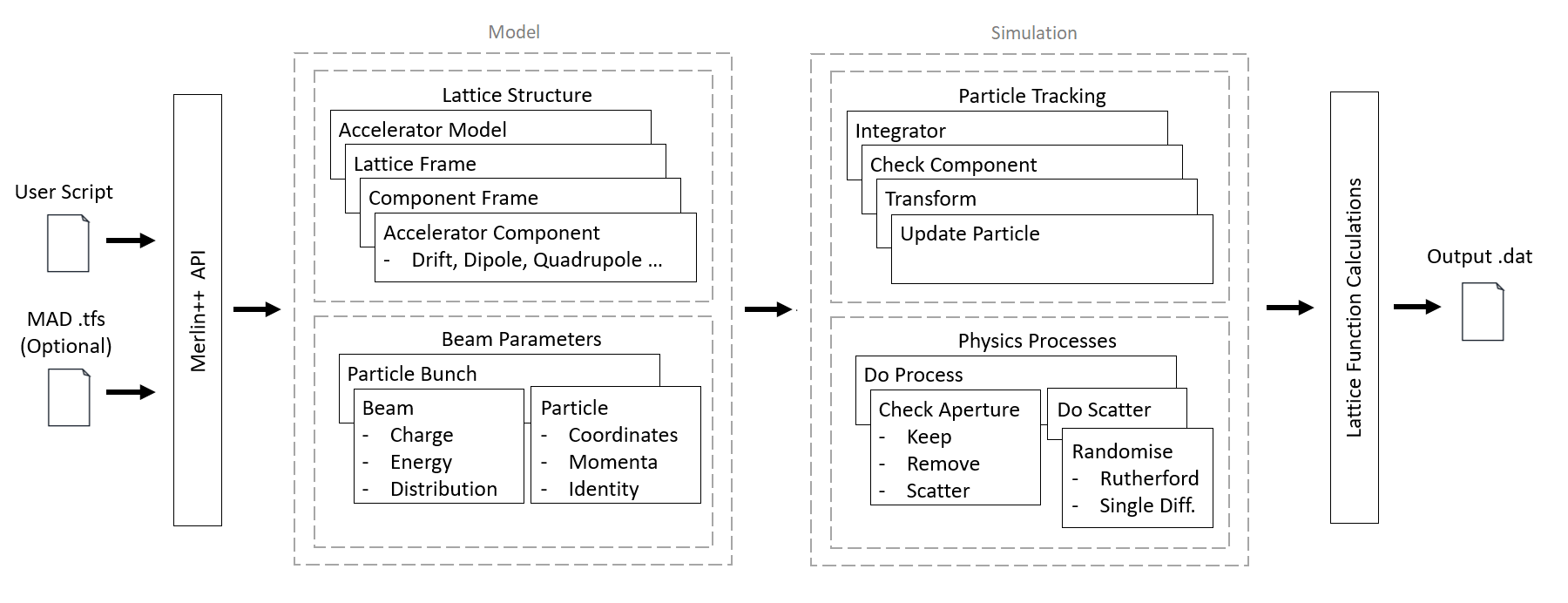}
\caption{\label{fig:architecture} High-level architectural view of Merlin++ lattice construction and particle bunch tracking.}
\end{figure*}

\noindent A graphical depiction of a simple particle tracking simulation based on the Merlin++ architecture is shown in Figure~\ref{fig:architecture}. The user program defines how the accelerator model and beam are created, and then calls the tracker to transport the particles through the lattice, applying requested physics processes at each stage.

To ensure a maintainable, modular and performant architecture, the C++ language was chosen as it contains powerful high-level OOD features such as templates and polymorphism, while also being able to manipulate low-level code to optimize program speed and memory access. Its is also a very widely taught programming language, with many physics and engineering students being familiar with it by the time they graduate. (Similar tracking codes written in FORTRAN77 and/or FORTRAN90 have a long history of usability and maintainability issues.) It was also decided that Merlin++ would remain a library rather than a stand-alone program to allow users to write use-case specific programs. This avoids the need to learn a domain-specific language. It does require
users to compile and link their program with the library before running but this is straightforward and is detailed in accompanying documentation. It can be compiled and run in a terminal window and/or within an IDE such as Eclipse CDT \cite{Eclipse}. Users can add new components and features and/or combine Merlin++ with custom or third party code, such as ROOT~\cite{ROOT}.  Merlin++ uses the CMake~\cite{cmake} build and test package for compilation 
and for running tests. 

\subsection{Lattice Structure}
The lattice structure can be either be defined by specific calls in a  user program, or imported from a MAD output file. The design follows the core concept of models,  frames, elements and components. A top-down perspective would see an {\tt AcceleratorModel} class being constructed to contain a {\tt LatticeFrame} - itself comprised of either contiguous sections or a complete {\tt Beamline} or {\tt Ring} (defined model types). A frame is a geometrical construct, a local coordinate frame, but one may consider it as a hollow structure which outlines the lattice structure. Each {\tt Beamline} or {\tt Ring} section contains within it a number of
{\tt ComponentFrame}s, providing a frame for each element. A {\tt ComponentFrame} hosts a {\tt ModelElement} which may or may not be a pre-defined {\tt AcceleratorComponent}, with aperture, electromagnetic field, and geometry properties. Component classes, {\tt SimpleDrift, SectorBend, Quadrupole} etc, are derived classes of {\tt AcceleratorComponent}. The generic {\tt ModelElement} root class allows users to define and seamlessly integrate additional use-case specific components such as power supplies, klystrons  etc. Tracking simulations use provided iterators to cycle through each {\tt ModelElement} to call location-specific {\tt AcceleratorComponent} properties.

\subsection{Beam Parameters}

Merlin++ has a generalised concept of particle bunches, with separation between high level parameters of a beam and how it is represented in a given simulation. The {\tt BeamData} class holds the parameters of the envelope of the beam at a given point (usually the start) of an accelerator, for example the centroid position, emittance and Courant-Snyder parameters. Typically these are set from the values obtained by the {\tt LatticeFunctionTable} as described in section~\ref{calc_lat_func}. From these parameters a representation of a bunch of particles is created. For example, a {\tt ParticleBunch} can be defined by a collection of {\tt Particle} phase-space coordinates, however, it is also possible to define the bunch by its moments using {\tt SMPBunch} (Sliced Macro-particle) class. If spin coordinates are required a {\tt SpinParticleBunch} can be used, as described in section~\ref{sec:spin-tracking}.

In a {\tt ParticleBunch} each {\tt Particle} is defined by a 6-D {\tt PSvector} (phase-space vector) of phase-space coordinates and momenta, as defined in \ref{appendixone}. These dynamical variables are updated at each step during tracking. 
By convention, the first particle in a bunch is placed on the reference orbit. Some physics processes use this reference particle for alignment.

\begin{figure} [htb]
  \centering
      \includegraphics[width=0.44\textwidth]{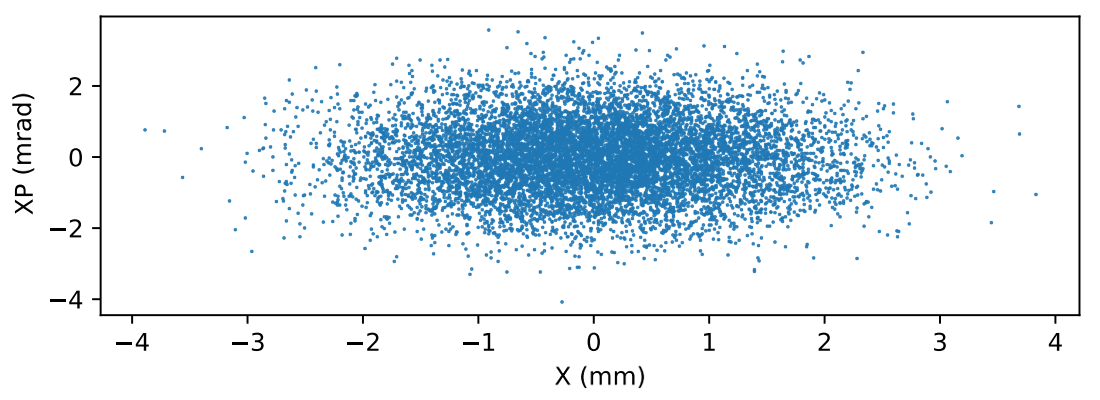}
      \includegraphics[width=0.44\textwidth]{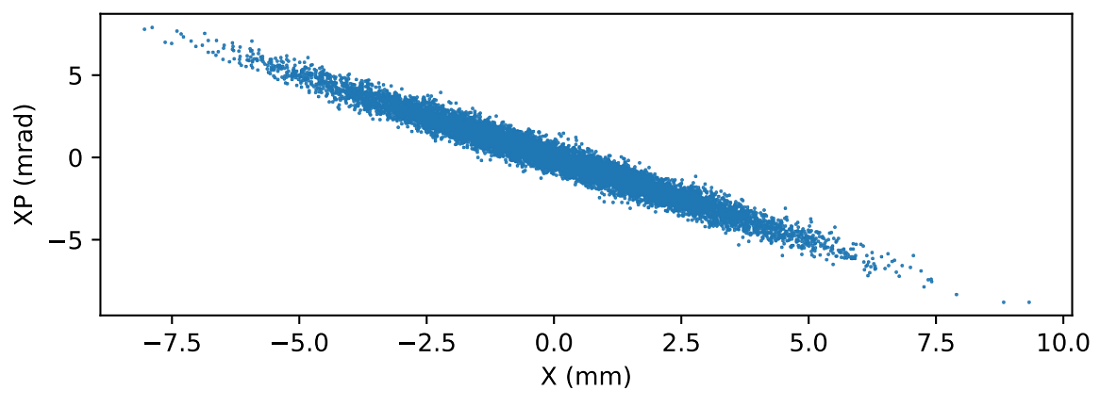}
      \includegraphics[width=0.44\textwidth]{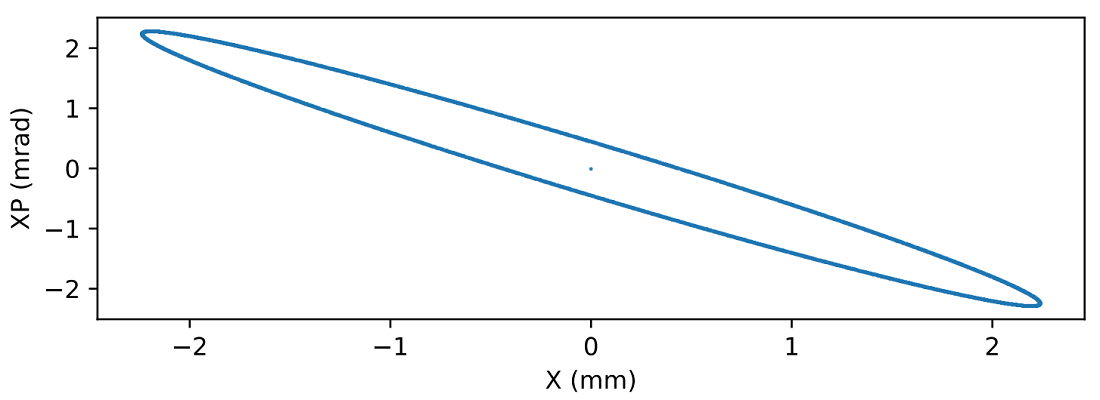}
  \caption{Bunch distribution profile in the $x$-$x'$ plane. Initially the bunch is created in normalized coordinates (top), then transformed by Courant-Snyder parameters (middle). A distribution from the halo option, set at one sigma,  is also shown  (bottom).}
\label{fig:bunch_dist}
\end{figure}

A {\tt ParticleBunch} can be created by several methods. An empty bunch can be created and particles then added to it with repeated calls to {\tt ParticleBunch::push\_back()}; this can be useful when very specific initial coordinates are required. Or the initial coordinates can be read from a file using {\tt ParticleBunch::Input()}, for example to import a bunch from another simulation code. Most typically a bunch with some standard distribution is required, and for this the {\tt ParticleBunch} constructor is called with a {\tt ParticleDistributionGenerator} and a {\tt BeamData}. Distribution generators for several common beam distributions including normal (Gaussian), flat and various halo models are available, and additional distributions can be added by the user. The distribution options originally create coordinates in a normalized space. Then, using beam parameters stored in a {\tt BeamData} object the coordinates are transformed by Courant-Snyder parameters to produce the final particle bunch. Phase-space distributions of normal and halo bunches are shown in Figure~\ref{fig:bunch_dist}.

\subsection{Calculation of Lattice Functions}
\label{calc_lat_func}
A common analysis problem is to calculate the  lattice functions, also known as the Courant-Snyder or Twiss parameters, which determine the stability of a periodic lattice and are also relevant for transfer lines.  Merlin++ implements a  normal form approach method of calculation following Wolski \cite{Wolski1,Wolski2}. The closed orbit is calculated as usual and the 1-turn matrix $\bf M$ is evaluated at an arbitrary point by tracking single particles with small displacements from the closed orbit. If the orbit is stable, then $\bf M$ has eigenvalues $e^{\pm i \mu_k}$, with $\mu_k$ real, where the index $k$ runs from 1 to 3 and can be identified, if the coupling is not too strong, with motion in the horizontal, vertical and longitudinal directions.
A related matrix, ${\bf \Sigma}_{ij}=\left< x_i x_j\right>$ which describes matched bunches ({\it i.e.} which do not change after a complete turn: $ {\bf M  \Sigma \tilde  M} = {\bf \Sigma}$) can be written as the sum of 
three matrices $\sum \epsilon_k \bf{B}_{\textit k} $, where $\epsilon_k$ are the emittances
and elements of the $6 \times 6$ matrices $\bf{B}_{\textit k}$ and map on to the lattice parameters in the absence of coupling. From the eigenvectors of $\bf M$, a matrix $\bf N$ can be constructed which transforms the general 6-D turn by turn motion into three normal 2-D circular forms, and the elements of $\bf N$ give the elements of $\bf B$. Subsequently, after each of the $\bf {B}_{\textit k}$ matrices have been determined for an arbitrary starting point, they can be tracked along the lattice and the corresponding lattice functions values can be extracted at each point. If there is no coupling then the Courant-Snyder functions can be obtained from the appropriate element of the appropriate matrix, {\em e.g.} (in form $\bf B^{\textit k}_{\textit i,\textit j}$) $\alpha_x=-B^1_{12}$, $\alpha_y=-B^2_{34}$, $\beta_x=B^1_{11}$, {\em etc}.
Dispersion is calculated by $\eta_x=B^3_{16}/B^3_{66}$ and $\eta_y=B^3_{36}/B^3_{66}$.
Merlin++ keeps track of these indices using instances of the {\tt LatticeFunction} class. If requested, values are stored in a {\tt LatticeFunctionTable}. Entries are requested by {\tt AddFunction(i,j,k)} method, where $i,j,k$ are the corresponding matrix elements. For simplicity, a {\tt UseDefaultFunctions} function is provided which calculates and stores 10 common functions: $x, p_x, y, p_y, ct, \delta, \beta_x, -\alpha_x, \beta_y, -\alpha_y$. These functions can then be readily plotted by a {\tt python} script, {\tt gnuplot} or a
similar plotting program. 

\subsection{Non-linearities and Perturbation}
\begin{figure} [h]
  \centering
      \includegraphics[width=0.237\textwidth]{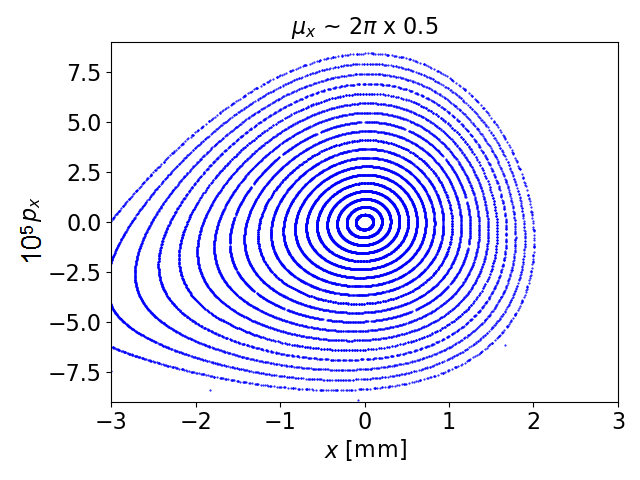}
      \includegraphics[width=0.237\textwidth]{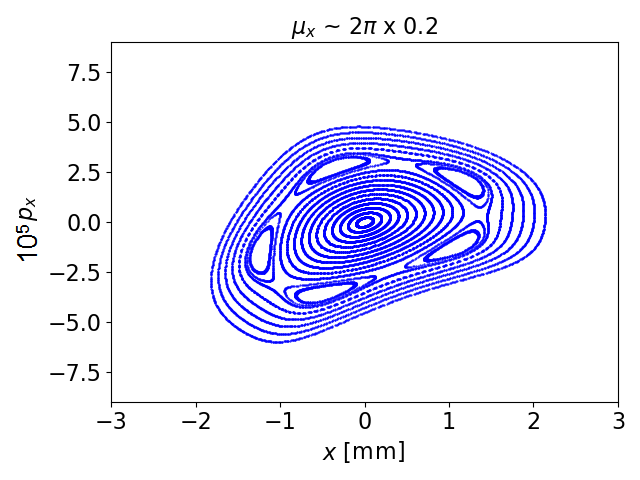}
      \includegraphics[width=0.237\textwidth]{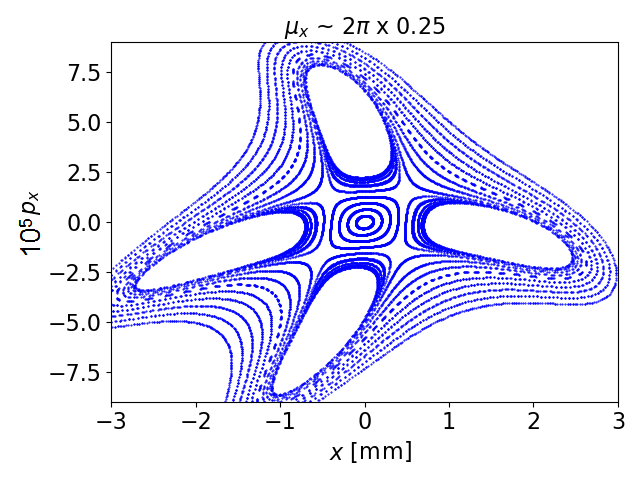}
      \includegraphics[width=0.237\textwidth]{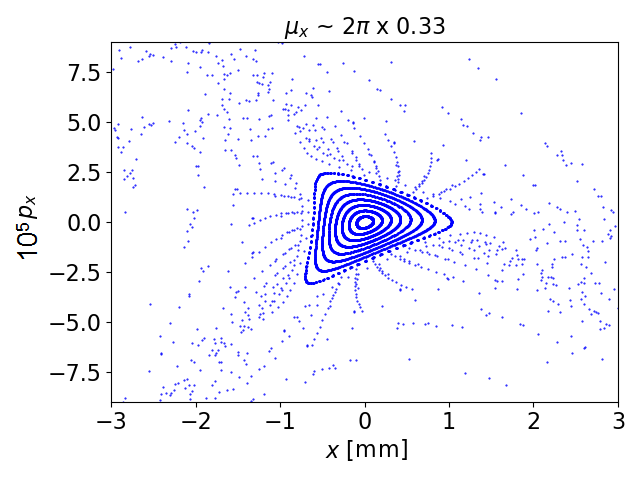}
  \caption{The effect on the horizontal parameters of adding a constant length/field sextupole to the end of a periodic FODO lattice cell. Shown are half, third, quarter and fifth integer phase advances $\mu_x$.}
\label{fig:sextupole_perturbation}
\end{figure}

In engineering operationally stable accelerators, one must take into account non-linearities and the resultant instabilities. Such effects may arise as a direct result of a high-order magnetic component or from collective dynamics  within a particle bunch. Utilizing Merlin++'s extensive lattice function capabilities, a user may isolate and monitor individual components or lattice regions to analyse non-linearities and the long-term affect on stability. For example, if a perturbation effect of magnetic components occurs when operating near a specific order-related phase advance 2$\pi$ integer fraction, repeated effects accumulate. Figure~\ref{fig:sextupole_perturbation} shows a simple example of adding a sextupole to the end of an otherwise stable periodic FODO lattice cell. As shown, varying the  phase advance (by adjusting the quadrupole strengths) can result in significant instabilities. Such analysis is important in optimizing the dynamic aperture.

\subsection{Particle Tracking Integrators}
Merlin++ provides various integrator options for particle tracking.   The {\tt SYMPLECTIC} integrator is the default and is what   advised for simulations of large numbers of turns. 
  {\tt TRANSPORT} and {\tt THIN\_LENS} are also available for bench-marking against other software packages. In each case a given particle, defined by its {\tt PSvector} variables is subject to an explicit transformation as it propagates through a component.




\subsubsection{The Symplectic Integrator}
Any numerical integration inevitably introduces inaccuracy.
A symplectic integrator preserves the Hamiltonian, though it is not necessarily the most accurate integrator for some individual particle parameter.

From the accelerator Hamiltonian, individual component integrators are derived by first defining the component specific electromagnetic fields and simplifying the Hamiltonian, accordingly. Using Hamilton's equations, one establishes and solves the equations of motions to arrive at a transfer map and/or explicit equations for each dynamical variable. Not all component Hamiltonians are integrable and are split into  drifts and kicks: $H = H_{d}+H_{k}$, where the drafts are independently integrable and the kicks requireapproximation,  commonly of 2\textsuperscript{nd}-order, and may use a  small angle (paraxial) approximation,

The loss in accuracy can be minimised by splitting the Hamiltonian further, {\em e.g.} $H = H_{d}/2+H_{k}+H_{d}/2$. This splitting method can be extrapolated to much higher-order \cite{yoshida} using Lie algebra manipulation. 
For the drift component, one may integrate 
the exact Hamiltonian. Furthermore, Forest has shown that it is possible to derive an exact solution for rectangular and sector bend dipole magnet components \cite{forestbook1}. This allows for a considerable increase in accuracy and 
use of the splitting method with the exact drift and sector bend components forms the basis of the Merlin++ symplectic integrator and  makes it a suitable 
software package for long-term 
tracking and beam-lifetime studies. 
Details are given in \ref{appendixtwo}.

\subsubsection{RF Cavity Structures}
Merlin++ has a number of RF options, including both travelling and standing wave structures. Simplified structures are also available which have no transverse focusing. Transverse magnetic mode (TM\textsubscript{010}) structures simulate acceleration by by change  in the particle momentum:
\begin{equation}
\Delta\delta = \Delta p_{z}-\Delta p_{z,\text{ref}} = - \frac{q V}{P_{0}c}[\cos(\phi+\Delta\phi)-\cos(\phi)].
\end{equation}
\noindent where $V$ is the cavity voltage and $\phi$ is the phase. This can be followed by a redefinition of the reference momentum to avoid excessive values of $\delta$.  Where required, transverse focusing is included, for the appropriate RF wave structure,  in accordance with Rosenzweig and Serafini \cite{Rosenzweig-Serafini}, such that 
\begin{equation}
\bf M =
\begin{pmatrix}
\cos(\alpha)- \sqrt{2}\cos(\phi)\sin(\alpha) & \frac{\gamma_{\text{i}}}{\gamma_{\text{f}}}\sqrt{8}\cos(\phi)\sin(\alpha)\\
-\frac{\gamma_{\text{i}}}{\gamma_{\text{f}}}(\frac{\cos(\phi)}{\sqrt{2}}+\frac{1}{\sqrt{8}\cos(\phi)})\sin(\alpha) & \frac{\gamma_{\text{i}}}{\gamma_{\text{f}}}\left(\cos(\alpha)+\sqrt{2}\cos(\phi)\sin(\alpha)\right)
\end{pmatrix},
\end{equation}

\noindent where $\bf M$ is the transfer map, $\gamma_{\text{i}}$ and $\gamma_{\text{f}}$ are the initial and final Lorentz factors and 
\begin{equation}
\alpha =\sqrt{\frac{1}{8}} \frac{1}{\cos(\phi)}\log{\left(\frac{\gamma_{\text{f}}}{\gamma_{\text{i}}}\right)}.
\end{equation}

\noindent A {\tt Klystron} class also allows configuration of arrays of cavities to model the RF control and power system.

In addition to the above, transverse electric (TE) mode structures have be implemented to simulate crab cavities for HL-LHC studies. The transverse kick follows
\begin{equation}
\Delta p_t  = -\frac{e V_{cc}}{E_0}\sin(\phi+\Delta\phi),
\end{equation}

\noindent where $V_{cc}$ is the strength of the cavity and $E_0$ the reference energy. The kick strength may be dynamically altered during simulation to model a cavity failure via the {\tt CCFailureProcess}.

\subsubsection{Sliced Macroparticle Tracking}
In situations where a user is simulating linear lattice models or does not require detailed bunch descriptions, a sliced macroparticle (SMP) model of a bunch may be used. For a {\tt SMPBunch}, phase-space coordinates are replaced by 1\textsuperscript{st} and 2\textsuperscript{nd} order {\em moments}. In this case, the normal tracking integrator does not work and corresponding {\tt SMPComponentTracker} and {\tt SMPStdIntegrators} classes define specific equivalents. The  SMP integrator does not fully model non-linear lattice elements, although it does include feed-down effects from them. SMP bunches allow for implementation of collective effect physics processes and have been extensively used in wakefield simulation studies.

\subsubsection{Spin Tracking}
\label{sec:spin-tracking}
Spin tracking was implemented to simulate electron beam depolarisation in the Next Linear Collider (NLC) Main Damping Rings \cite{merlin_nlcspin}. Electron spin dynamics in a high-energy storage ring can be described by the Thomas-BMT equation \cite{thomas, BMT}, where the precession of the spin vector, $\vec S$, follows
\begin{equation}
\frac{d\vec S}{dt} = \vec\Omega \times \vec S
\end{equation}

\noindent where $\vec\Omega$ is constructed from  the electromagnetic fields. Assuming the electric field is zero, $\vec\Omega$ can be defined, for linear coordinates, by 
\begin{equation}
\vec\Omega = -\frac{e}{\gamma m}[(1+G\gamma)\vec B_{\bot}+(1+G)\vec B_{\parallel}]
\end{equation}

\noindent and for curvilinear coordinates, by 
\begin{equation}
\vec\Omega = -\frac{e}{\gamma m}[G\gamma\vec B_{\bot}+(1+G)\vec B_{\parallel}]
\end{equation}

\noindent where $G$ is the anomalous magnetic moment (0.00115965 for electrons) and $\vec B_{\bot}$ and $\vec B_{\parallel}$ are perpendicular and parallel electron motions, respectively. In addition to tracking the above, Merlin++ also implements dipole magnet fringe fields for spin tracking as they can change the vertical component of the spin vector. The fringe field model used  is  
\begin{equation}
\int B_{z}ds = \frac{1}{2}y\hat B_{y}.
\end{equation}

\subsubsection{Physical Aperture Boundaries}
A powerful tool of the Merlin++ tracking routine is physical aperture boundary checking, if a {\tt CollimateParticleProcess}
 is activated. If aperture information is provided, following every discrete propagation along a beamline, a particle's coordinates are compared against the physical aperture boundaries and if it exceeds these the particle is considered lost or, if the  {\tt CollimateProtonProcess} child of {\tt CollimateParticleProcess} is used,  be subject to a physics process, such as {\tt ScatteringProcess}. Merlin++ supports a number of aperture geometries, including {\tt Rectangular}, {\tt Circular}, {\tt Elliptical}, {\tt Rectellipse} and {\tt Octagon}. If additional aperture shapes are required the user can create a subclass of  {\tt Aperture} and override the {\tt CheckWithinApertureBoundaries} method. Element-specific step sizes can be defined, for example, every 50~cm or only at the beginning and end of an element. If more accurate checks are required, {\em e.g.} for a collimator, and the provided aperture data is less discrete than the step size, linear interpolation of aperture data is carried out prior to boundary checks.

\subsection{Physics Processes}
In many cases a user may wish to add additional physics processes for the tracker to call, {\em e.g.} collimation, synchrotron radiation {\em etc}. Physics processes are defined in the {\tt BunchProcess} class. When a process is activated for a given simulation, the tracker will call the process at each model element. At this point, the process will determine if the physics routine should be run, for example the synchrotron radiation process may be run for a dipole, but not a drift component. 
A priority list is defined when multiple processes are activated such that dominant processes are called first. A number of commonly used physics processes are detailed in the follow subsections. Note, however, that processes are defined generically and are flexible in nature. As such, they may used for practical purposes beyond physics functions. For example, the {\tt MonitorProcess} class allows the user to read and output a particle's coordinates as it traverses a set of elements.

\subsubsection{Synchrotron Radiation}

The loss in particle energy due to being accelerated in a curved path or radially with respective to the current trajectory  is known as synchrotron radiation. Following Burkhardt \cite{synchradmc}, the {\tt SynchRadParticleProcess} implements a Monte Carlo photon spectrum generator. The total number of photons and/or the total energy loss is derived from the following relation in accordance with Schwinger \cite{synchradclassic},
\begin{equation}
\int_{0}^{\infty} x^n K_v(x)dx = 2^{2n-1}\Gamma\Bigg(\frac{1+n-v}{2}\Bigg)\Gamma \Bigg(\frac{1+n+v}{2}\Bigg),
\end{equation}

\noindent where $n$ is the number of radiated photons, $v$ the particle velocity and $K_v$ is a modified Bessel function. The $\Gamma$ function has the useful property, $\Gamma(x)\Gamma(1-x)=\pi/\sin(\pi x)$. Merlin++ allows for the loss of energy to be taken into account by two different methods: first, by calculation and modification of individual particle momenta, leaving the reference momentum unchanged; second, by determination of the mean energy loss, which is then applied to the reference momentum and individual particles, accordingly. 
Merlin++ does not currently track the produced photons, although an implementation would be possible.

\subsubsection{Collimation and Scattering} \label{sec:collimation}
Controlled removal of outer particles in unstable orbits is common practice in accelerator operations. This process of  active collimation involves installation of high energy-capacity collimators with adjustable mechanical jaws designed to actively reduce the local physical aperture and absorb  particles that would otherwise be lost in more vulnerable components. Collimators are typically installed at high $\beta$ locations along a lattice structure.

At high beam energies the collimator will not absorb the complete energy of a particle, and of any secondary particles it may produce. Systems of primary (spoiler) and secondary (absorber) collimators are used. The simulation of such complete systems requires detailed modelling, and is done by programs such as FLUKA~\cite{FLUKA1,FLUKA2} and Geant4~\cite{Geant4} (including BDSIM~\cite{BDSIM}) at a level of detail which is not compatible with a program for simulating accelerator optics, and Merlin++ does not attempt to provide a
fine-grained description of such showering processes. However it is appropriate for an accelerator optics simulation to consider particles which undergo a single scatter in a collimator, experiencing a small deflection and loss of energy, and continue to pass though the lattice, eventually being lost at some remote location -- indeed, perhaps after experiencing several turns through the lattice.

The {\tt Collimator} class provides functionality to do this, being able to model collimators with various geometries (including a  one-sided jaw), and/or alignment errors, and also specific material properties, including composite materials, as described in section~\ref{sec:materials}. 

The {\tt CollimateParticleProcess} implements a series of checks to determine if a particle should be lost and/or scattered. While propagating through a {\tt Collimator} model element, if a particle is found to exceed adjusted physical aperture boundaries, but remains within the original beam-pipe boundaries, the particle is considered   scattered.
To model such an event, the {\tt ScatteringProcess} class implements a number of possible scattering options, selected randomly according to their probability.
Absorption (which is taken as including inelastic scattering and the start of a particle shower) is one of these; others are elastic and quasi-elastic (diffractive) scattering.
Scattering options include common models for multiple coulomb, inelastic proton-nucleus, and Rutherford scattering. Moreover, Merlin++ implements uniquely developed models for elastic proton-nucleus and single diffractive proton-nucleon scattering which covers the full kinematical range of LHC energies.
Full details are given in \cite{pomeronpaper}.

The model for elastic and single diffractive scattering incorporates both Coulomb and nuclear amplitudes as well as the interference between the two. For elastic scattering, four Regge trajectories, the hard Pomeron, soft Pomeron, the $f_2$ and $a_2$ trajectory and the $\omega$ and $\rho$ trajectory are used for both Coulomb and nuclear amplitudes. For nuclear amplitudes an additional triple-gluon exchange is included. 
The full model of the differential cross-section for elastic scattering follows 
\begin{multline}
\frac{d\sigma}{dt}=\pi[A_c(s,t)]^2+\frac{1}{4\pi}(\Re [A_n(s,t)]^2+ \\\Im [A_n(s,t)]^2)+(\rho +\alpha_{\text em}\phi)A_c(s,t)\Im [A_n(s,t)], 
\end{multline}

\noindent where $\sigma$ is the interaction cross-section, $s$ and $t$ are invariants of motion, $\alpha_{\text em}$ the Regge trajectory, $\phi$ the Coloumb phase, $\rho$ the ratio of real and imaginary components of the nuclear term and $A_c$ and $A_n$ are Coulomb and nuclear amplitudes, respectively.

The single diffractive model follows the triple-Regge description of high-mass diffractive dissociation. The implemented double differential cross section for high missing mass is
\begin{equation}
\begin{split}
\frac{\partial^2\sigma^{\text{HM}}}{\partial t\partial \xi}(\xi,t,s)  = & g_{\mathbb{P}\mathbb{P}\mathbb{P}}(t)s^{\alpha_{\mathbb{P}}(0)-1}\xi^{\alpha_{\mathbb{P}}(0)-2\alpha_{\mathbb{P}}(t)} + \\
& g_{\mathbb{P}\mathbb{P}\mathbb{R}}(t)s^{\alpha_{\mathbb{R}}(0)-1}\xi^{\alpha_{\mathbb{R}}(0)-2\alpha_{\mathbb{P}}(t)} + \\
& g_{\mathbb{R}\mathbb{R}\mathbb{P}}(t)s^{\alpha_{\mathbb{P}}(0)-1}\xi^{\alpha_{\mathbb{P}}(0)-2\alpha_{\mathbb{R}}(t)} + \\
& g_{\mathbb{R}\mathbb{R}\mathbb{R}}(t)s^{\alpha_{\mathbb{R}}(0)-1}\xi^{\alpha_{\mathbb{R}}(0)-2\alpha_{\mathbb{R}}(t)} + \\
&\frac{g^2_{\pi\pi p}}{16 \pi^2}\frac{|t|}{(t-m_{\pi})^2}F^2(t)\xi^{1-2\alpha_{\pi}}(t)\sigma_{\pi^0 p}(s\xi),
\end{split}
\end{equation}
\noindent where $\xi=M^2_{\text X}/s$ with $M_{\text X}$ the missing mass,
$\mathbb{P}$ and $\mathbb{R}$ denote Pomeron and effective Regge trajectories, respectively, and the 5\textsuperscript{th} component is the pion exchange term. For low missing mass interactions, appropriate resonances are included and the differential cross section is given by
\begin{equation}
\frac{\partial^2\sigma^{\text{LM}}}{\partial t\partial \xi}(\xi,t,s) = \frac{\partial^2\sigma^{\text{HM}}}{\partial t\partial \xi}(\xi,t,s) + \frac{\partial^2\sigma^{\text{res}}}{\partial t\partial \xi}(\xi,t,s) + R_{\text m}(\xi,t,s),
\end{equation}

\noindent $\sigma_{\text{res}}$ is the resonance region cross section and $R_{\text m}$ is a resonance matching term to account for a small step when $\xi = \xi_{\text c}$.

\subsubsection{Hollow Electron Lens}

In some cases, relatively tight collimation is desired, but must be done without significantly increasing beam impedance. A proposed solution for proton beam accelerators is the use of so-called hollow electron lenses (HEL) -- a hollow cylindrical distribution of electrons aligned in the transverse plane to create a ring for a beam to pass through. $R_{min}$ and $R_{max}$  give the inner and outer radii of the cylinder. For a perfect cylindrical symmetric lens, inside the hollow region ($r < R_{min}$) there is no net field produced, so the core of the proton beam is unaffected. A proton at above $R_{min}$ will feel the electric and magnetic components of the electrons at lower radii. 

For opposite charges (such as an electron lens with a proton beam), the HEL deflects high amplitude particles radially inwards towards the lens centre. The cumulative effect over many turns is to increase the amplitude of these particles, driving them onto the primary collimators. This allows for more relaxed physical aperture dimensions and an overall reduced impedance. Functionality to model HELs in Merlin++ has been implemented following Rafique~\cite{rafique_thesis_2017}. Assuming a radially symmetric electron distribution, the kick applied to a proton at radial position $r$ within the HEL radial boundaries is given by
\begin{equation}
\theta(r)= f(r) \frac{1}{4\pi\epsilon_0c^2}\frac{2LI(1\pm\beta_e\beta_p)}{(B\rho)_p\beta_e\beta_p}\frac{1}{r},
\end{equation}

\noindent where $L$ is the HEL length, $I$ the electron beam current, $(B\rho)_p$ the proton beam rigidity and $\beta_e$ and $\beta_p$ are Lorentz $\beta$ factors for the electron and proton beams, respectively. The $\pm\beta_e\beta_p$ is positive if the beams propagate in opposite directions. The $f(r)$ gives the fraction of electrons enclosed at a given radius, for a uniform electron distribution it is given by:
\begin{equation}
f(r) =
\begin{cases}
	0   & \text{if } r < R_{min}\\
	\frac{r^2 - R_{min}^2}{R_{max}^2 - R_{min}^2} & \text{if } R_{min} < r < R_{max} \\
	1   & \text{if } R_{max} < r
\end{cases}
\end{equation}

\subsubsection{Wakefields}
Merlin++ contains the ability to calculate effects of intra-bunch wakefields, both geometric wakefields produced by changes in beam pipe aperture \cite{merlinwakefield1} and resistive wakefields due to induced currents~\cite{merlinwakefield2}.

A particle at radial position ($r'$,$\theta'$) induces currents in the beam pipe whose electric and magnetic fields affect a later particle at position ($r$,$\theta$). These wakefield effects alter the shape of the bunch, and in some cases this can result in a serious increase in emittance. 
For a perfectly conducting beam pipe of constant aperture there is no wakefield effect: wakefields due to varying geometry or finite resistivity are treated separately.

In both cases
wakefields are calculated using an expansion in angular modes. The {\tt SMPBunch} (Sliced MacroParticle) version of {\tt ParticleBunch}  sorts the particles into longitudinal order - this need only be done once.
For a given mode $m$, and defined moments $C_j^m=\sum r'^m\cos(m\theta')$ and $S_j^m=\sum r'^m\sin(m\theta')$, the combined kick (momentum imparted) to a particle in slice $i$ trailing $N_j$ particles in slice $j\geq i$ by a distance $s_{ji}$ is
\begin{equation}
W_x=\sum_mmr^{m-1}\left(\cos(\mu\theta)\sum_jW_m(s_{ji})C_j^m+\sin(\mu\theta)\sum_jW_m(s_{ji})S_j^m\right)
\end{equation}
\begin{equation}
W_y=\sum_mmr^{m-1}\left(-\sin(\mu\theta)\sum_jW_m(s_{ji})C_j^m+\cos(\mu\theta)\sum_jW_m(s_{ji})S_j^m\right)
\end{equation}
\begin{equation}
W_\parallel=\sum_m W'_m(s_{ji})r^{m}\sum_j\left(S_j^m\cos(m\theta)+C_j^m\sin(m\theta)\right)
\end{equation}

\noindent which uses the wake functions $W_m$ for a given aperture geometry. These (not to be confused with the wake fields themselves) describe the effect of one particle on a later particle in a particular geometry. For example, for a perfectly conducting tapered circular collimator, with radius dimension tapering from $a$ to $b$, the wake function is
\begin{equation}
W_m(s)=2\Bigg(\frac{1}{a^{2m}}-\frac{1}{b^{2m}}\Bigg)e^{-ms/a}\Theta(s),
\end{equation}
\noindent where $s$ is the longitudinal interval between leading and trailing particles and  $\Theta(s)$ is a unit step function (ensuring later slices do not affect earlier ones).

In Merlin++ the wakefield computation and its effect can be found to any order $m$ - unlike most wakefield simulation programs which only consider the dipole term. The downside of this is that such wake function calculations and their implementation have only been done for circular apertures.
Calculations for other shapes - of which parallel-jawed collimators are typical~\cite{merlinwakefield3} - can only be indicative.

\subsection{Use Cases Examples}
Merlin++, has a relatively long development history and has been used to model and study  phenomena in a wide array of accelerators. Additional functionality has often been added to support these studies. Reviewing these studies gives an overview of the features and scope of the Merlin++ code base, and the extensibility of the core code design. 

\subsubsection{ILC Studies}
Merlin++ was originally developed at DESY to support the design and optimization of the~TESLA~\cite{schulte_simulations_2003} and ILC electron linacs. In addition, a stand-alone package called ILCDFS was created using the Merlin++ library, to simulate luminosity stability and model methods of dispersion free steering (DFS)~\cite{merlin_ilcdfs}. ILCDFS implements a beam-based alignment technique to minimise emittance growth within in a linac. This application could be constructed using Merlin++ as it is possible to first define a reference lattice and then modified versions to take into account misalignments and magnet errors. ILCDFS also models ILC collimator wakefields, alignment errors and dynamic ground motion~\cite{merlin_ilcground}. Moreover, the generic {\tt ROChannel} and {\tt RWChannel} classes were used to develop an iterative correction algorithm {\tt DFSCorrection}. ILCDFS is provided as a stand-alone application example with the Merlin++ source code.

\subsubsection{NLC Studies}

Merlin++ was used in the development of the main damping rings of the NLC, which needed to preserve the 80\% beam polarization provided by the source. As the program included spin tracking, as described in section \ref{sec:spin-tracking}, it could be used for the tracking studies necessary to complement semi-analytic calculations of the effects of spin resonances, including magnet misalignments and fringe field effects.

These
 showed that the nominal operating energy was safe from strong resonant effects, but suggested the design should allow for small energy adjustments in case a weaker resonance caused unexpected issues~\cite{merlin_nlcspin}.

\subsubsection{LHC Studies}
  Merlin++ has been used to support LHC and HL-LHC collimation projects~\cite{molson_proton_2014,tygier_performance_2019}. The LHC is a 27~km, 7~TeV proton synchrotron. Due to the large stored energy in the beam, it is possible that scattered or lost particles can deposit 
  sufficient energy in the super-conducting dipoles to quench the magnets, so beam losses around the ring must be controlled. 
  Movable collimators at locations around the LHC ring can be adjusted to clean the bunch halo in a safe manner. At the high energies used, not all protons are absorbed by the collimators and many scatter and are eventually lost elsewhere. Merlin++ can simulate LHC collimation, 
  being able to install and move/rotate/correct model elements at any location, as well as construct and append new physics processes such as {\tt CollimateParticleProcess} and {\tt ScatterProcess}, detailed in Section~\ref{sec:collimation}. For the purposes of analysis, the final locations of lost particles are stored and loss maps can be generated. An example of a LHC collimation and loss map study is provided as an example/tutorial with the source code.

Merlin++ has been used to study many operational and planned configurations of the LHC. Examples include investigation of the effect of scattering models on losses in the cold regions~\cite{molson_proton_2014, pomeronpaper}, validation of losses during the dynamic changes of optics when  the beam the beams are `squeezed' for collision, and future configurations for HL-LHC~\cite{tygier_performance_2019}, and novel collimation schemes such as the Hollow Electron Lens~\cite{rafique_thesis_2017}.

\section{Software-Specific Features}
\label{sec:software}
\subsection{Coding Practices}
\begin{figure}[h]
  \includegraphics[width=0.48\textwidth]{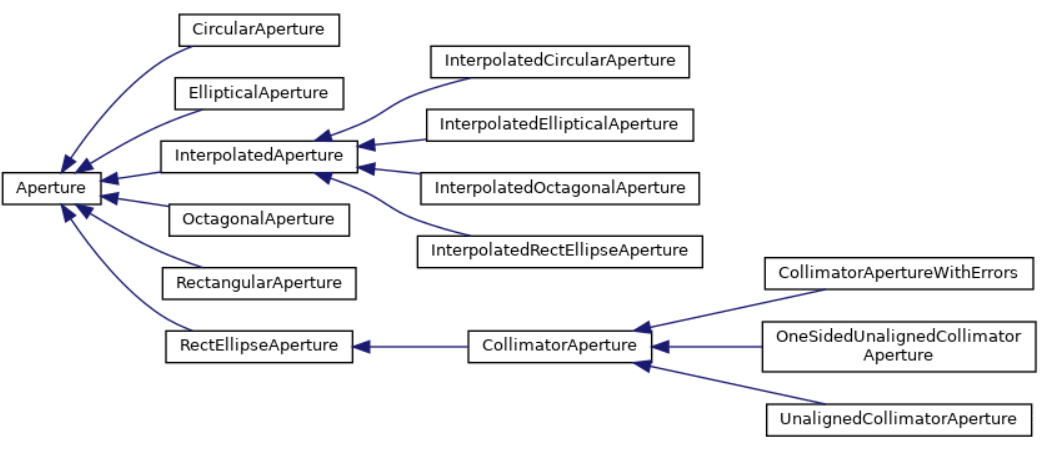}
  \caption{Example of a class inheritance structure within Merlin++, in this case the {\tt Aperture} class. The graphic is automatically produced by doxygen to be included in the documentation.}
\label{fig:apertureclass}
\end{figure}

Constructed in C++, powerful OOD practices are fundamental to the design and functionality of Merlin++. The original design philosophy was for Merlin++ to consist of a small number of loosely coupled, self-contained modules. This has been achieved by extensive use of inheritance and  polymorphism.
The IS-A/HAS-A principle and S.O.L.I.D~\cite{solid} design practices have been applied to all class structures. Moreover, all classes, methods and member function prototypes are accompanied by comments formatted to be read by the class library documentation generator Doxygen \cite{doxygen}, so users can generate Merlin++ class library documentation by use of the {\tt make doxygen} command. The class library is also available at the software website \href{http://www.accelerators.manchester.ac.uk/merlin/doxygen/}{/merlin/doxygen}{}. As an example the {\tt Aperture} class 
inheritance structure produced by doxygen is shown in Figure~\ref{fig:apertureclass}. 

Generic programming via polymorphic templates is also heavily used throughout the code base and has proven vital in the usability and maintainability of Merlin++ throughout its development. For example, the template class {\tt Matrix} is defined in {\tt TMatrixLib.h} where access methods are defined just once. A subsequent level of abstraction is then achieved by having specific matrix types defined by {\tt typedef Matrix<double> RealMatrix} such that developers can alter/add features more intuitively while the underlying core code remains untouched\cite{Barton}.

\subsection{I/O Channels}
A user will want specific component properties to be stored and/or dynamically altered during simulation. For this purpose, in addition 
to the standard library I/O, Merlin++ implements generic read-only (RO) and read-write (RW) {\em channels}. These are useful for
iterative tuning of orbit correction algorithms. For example, the {\tt ROChannel} may be used to read and store the output of a 
series of beam position monitors (BPMs) along a beamline. The collected data could then be used to calculate chromatic effects and the {\tt RWChannel} could then be scripted to alter sextupole corrector magnets, accordingly. This can 
be carried out iteratively within a loop until chromaticity errors are within desired limits. The user may then also re-use the {\tt ROChannel} to read and store sextupole component properties for future reference. 

\subsection{Utility functions}

Like any large program Merlin++ has numerous places that need similar small pieces of functionality, such as random numbers, linear algebra, file access and string matching. As a design goal, hard dependencies on external libraries have been avoided, so where functions are missing from the C++ standard library they have been implemented within the Merlin++ codebase.

An example is the {\tt DataTable} class inspired by structured array types in high-level languages such as the Python's numpy library~\cite{numpy}. A {\tt DataTable} is designed to hold tables of data with named columns that may have different types and provides accesses and modification as well as file input and output. This is especially useful for reading the TFS format used in MAD-X, for example for lattice and aperture descriptions via the {\tt MADInterface} class. Use of {\tt DataTable} has reduced duplication of file parsing code in several areas of Merlin++, while improving reliability.

In some cases additions to the C++ standard library have allowed removal of utility functions from Merlin++. High quality 
random numbers are needed in various parts of Merlin++ including bunch creation, element misalignment, scattering, synchrotron radiation and some modes of the hollow electron lens. C++11 added a random number library, which implements the Mersenne
Twister~\cite{matsumoto_1998_mersenne} algorithm and a number of distributions. Mersenne Twister is widely used in Monte Carlo simulations as it has a long period, high uniformity, low correlation between values up to high dimensions and passes many statistical tests for randomness. 
Switching to this standard implementation allowed
removal of several large source files from Merlin++ and solved some their deficiencies (slow warm-up time, limited seed space and less well studied generator function). 

\subsection{Code Testing}

Merlin++ features a suite of tests that can be run to confirm and verify correct building and functioning of the software. The tests can be run during development to prevent regressions in behaviour and quickly catch new bugs. 
The test suite is run automatically on each proposed change uploaded to the GitHub development platform, to confirm that the change can be successfully merged, built and run on Linux, Mac and Windows systems. In addition the tests are run on a variety of operating systems and architectures by an automated system after every commit to master, and every night.

Nightly tests are run using the CDash framework which is distributed with CMake. This provides a web interface showing recent test results. Dynamic analysis using Valgrind is also performed to flag issues such as uninitialised memory and memory leaks.

Tests include a mix of small unit tests for specific functions and larger tests of complete simulations. We are moving towards a test driven development model for new features to ensure additions function as expected.


\subsection{Developer Policy and Sustainability}
Software sustainability is described by Venters \textit{et al.} as `\textit{a software package’s capacity to endure in changing environments}' \cite{venters}. Today, this is achievable through strict adherence to modern software engineering practices, such as those outlined by the UKs Software Sustainability Institute (SSI) on the topics of usability and maintainability \cite{ssi}.

Merlin++ has had a number of core developers over the years and has included people from a range of backgrounds, each with varying levels of OOD and software engineering knowledge. As such, there was a notable disparity in code quality. To ensure that it is practical for Merlin++ to be utilized for years to come, current developers have invested significant time into profiling code quality metrics and reforming/refactoring code where necessary to re-establishing a consistent level of quality \cite{merlinsustainability}.

Quantitative analysis, carried out with Metriculator \cite{metriculator}, Valgrind \cite{valgrind} and ArchDia(R) DV8 \cite{dv8}, found a number of code complexity and dependency issues due to misuse or absence of appropriate OOD. Such issues limit readability and evolveability. Each issue was individually dealt with, either by simple refactoring and reformation to adherence to S.O.L.I.D. class design or a complete redesign, including implementation of appropriate design-patterns, {\em e.g.} leverage of decorated factory patterns for run-time type assignment to reduce the complexity of identified `God methods' containing an abundance of if/switch statements or similar. To prevent future occurrences of quality inconsistencies a number of developer policies were changed and strictly defined. Global formatting -- Uncrustify~\cite{uncrustify} -- and licensing -- GPL2+ \cite{gpl2} -- changes were applied, throughout. Merlin++ is now also on the stable repository platform, GitHub \cite{github}, at \href{https://github.com/Merlin-Collaboration}{github.com/Merlin-Collaboration}. In addition, all code being committed to the baseline must now be reviewed and approved by at least one other developer. A developer's guide detailing the above and more is now provided with the source code.

\subsection{C++14 and Beyond...}
As C++ has been developed with successive versions over the years, our design policy has been to learn and apply new functional improvements wherever they appropriately fulfill a design purpose or simplify code.
The Merlin++ 5.02 release introduced use of C++11 features, including smart pointers, lambda functions, move semantics and variadic templates.
The current development branch has begun to take advantage of features from C++14 and there is investigation into the concurrency potential of the C++17 standard library parallelism for future releases.
However, due to the timescale over which the code has been developed, it would be impractical to update/redesign a lot of the core legacy code structures for relatively minor performance improvements, unless profiling metrics point to  performance restrictions which could be circumvented by modern implementations. 

\section{Model Accuracy}

\label{sec:accuracy}

Merlin++ has been compared and benchmarked against a variety of related simulation packages\cite{Schulte:2002,Latina:2007} and to data taken from real accelerators. These studies give us confidence in its outputs. Below we give some examples of these comparisons.

\subsection{Tracking and Optics}

\begin{figure}[htb]
  \includegraphics[width=0.48\textwidth]{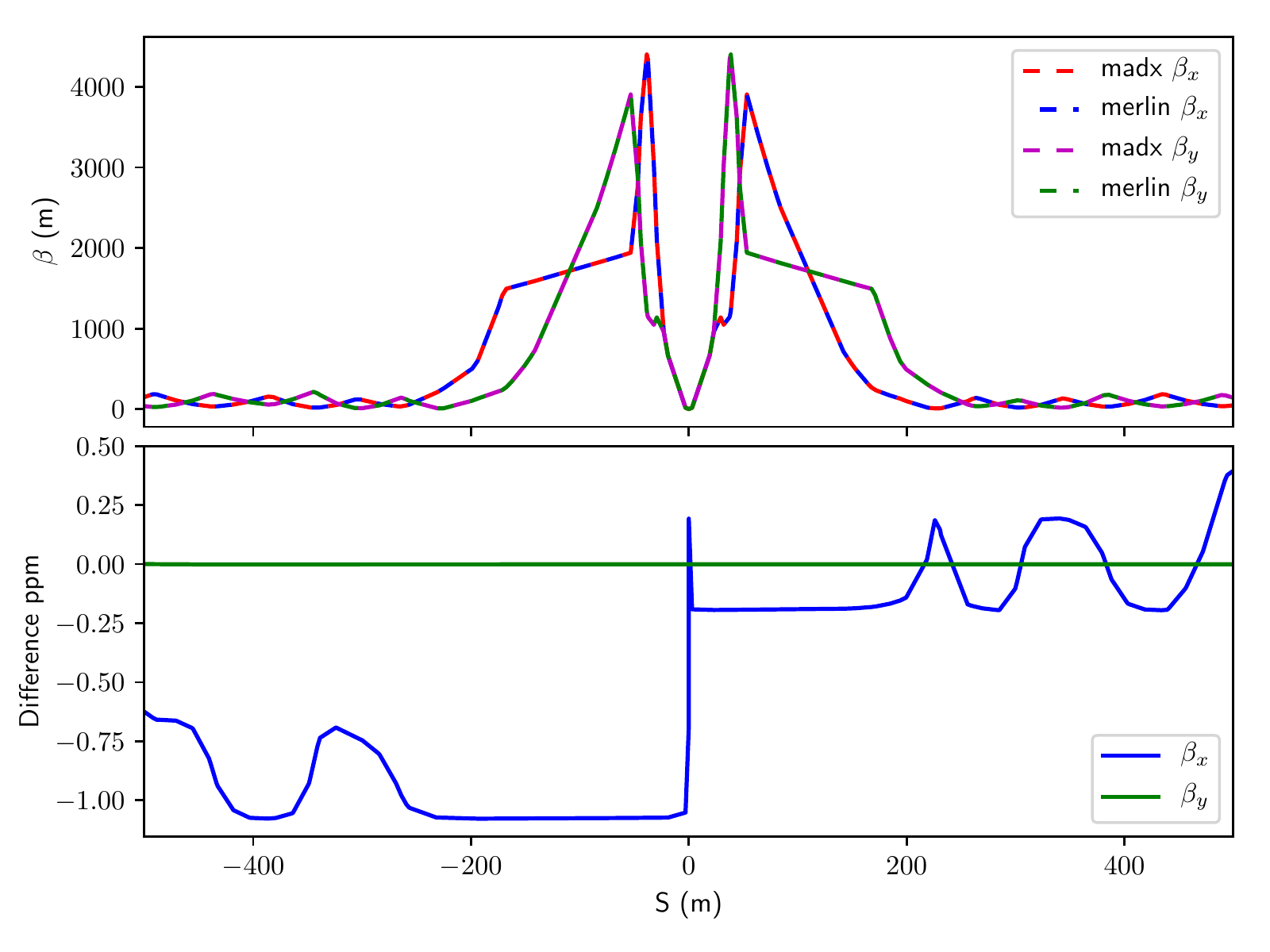}
  \caption{Comparison of $\beta$ functions around the ATLAS interaction region in the LHC. The lower plot shows difference between MADX and Merlin++'s output in parts per million.}
\label{fig:compare_optics_ATLAS_IR1_beta}
\end{figure}

MADX is widely used for design and optimization in the field of accelerator physics. It computes the optical functions of a lattice using the matrix formulation of each element, therefore providing a strong independent verification of Merlin++'s method using particle tracks. Figure \ref{fig:compare_optics_ATLAS_IR1_beta} shows the horizontal and vertical $\beta$ functions around the ATLAS interaction region in the LHC, comparing Merlin++ with MADX. The discrepancy is kept to around 1 part per million or lower even in this extreme optical configuration.

\subsection{Materials and Scattering}
\begin{figure}[htb]
  \includegraphics[width=0.48\textwidth]{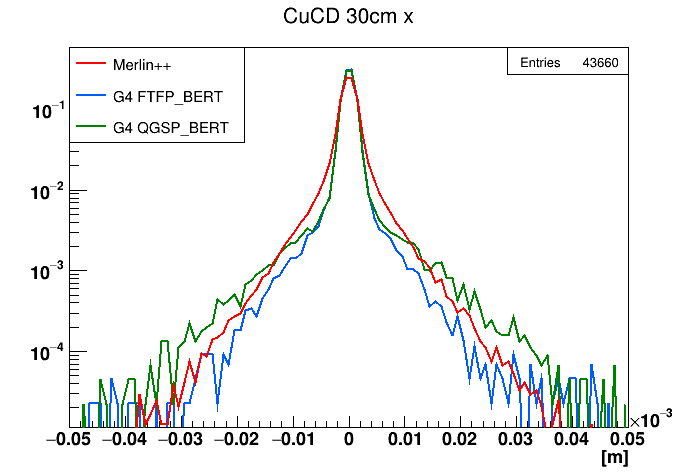}
  \caption{Comparison of fixed target horizontal scattering results between Merlin++ and Geant4. Geant4 physics lists used: QGSP\_BERT and FTFP\_BERT. Material used: copper-diamond (CuCD). Material depth: 30~cm.}
\label{fig:compare_geant4}
\end{figure}

Geant4 version 10.5~\cite{Geant4} 
was used to compare the implementation of Merlin++ scattering. A simple fixed target simulation was carried out in both Geant4 and Merlin++ with various materials of various depths. Two widely-used high-energy   physics lists were used to gain understanding of error boundaries in the theory: QGSP\_BERT and FTFP\_BERT. QGSP and FTFP refer to quark-gluon-string and Fritiof string excitation models for high energy interactions, above 10~GeV, respectively. BERT refers to the Bertini cascade model for low energy particles, below 10~GeV.
Figure~\ref{fig:compare_geant4} shows the comparative result for a copper-diamond (CuCD) collimator of 30~cm depth. It is clear that although simpler in nature, the Merlin++ scattering model yields similar results when compared to reputable Geant4 physics list options.
The central peak -- which includes most events -- is identical to the eye: at large angles the difference between the two Geant models is bigger than the difference of either with the Merlin++ curve. 

\subsection{Collimation and Loss Maps}
The LHC is highly instrumented to monitor beam losses around the ring, with about 4000 ionization chambers making up the Beam Loss Monitor (BLM) system. In addition to their role of triggering beam dumps if loss thresholds are exceeded they log losses during all machine operation.

Merlin++ is used in the LHC collimation group for simulating these proton losses, taking into account detailed scattering models in the collimators and the aperture limits around the entire ring.
These have been extensively compared to other simulations as well as to data taken during LHC operation~\cite{tygier_performance_2019}.

\begin{figure}[htb]
  \includegraphics[width=0.48\textwidth]{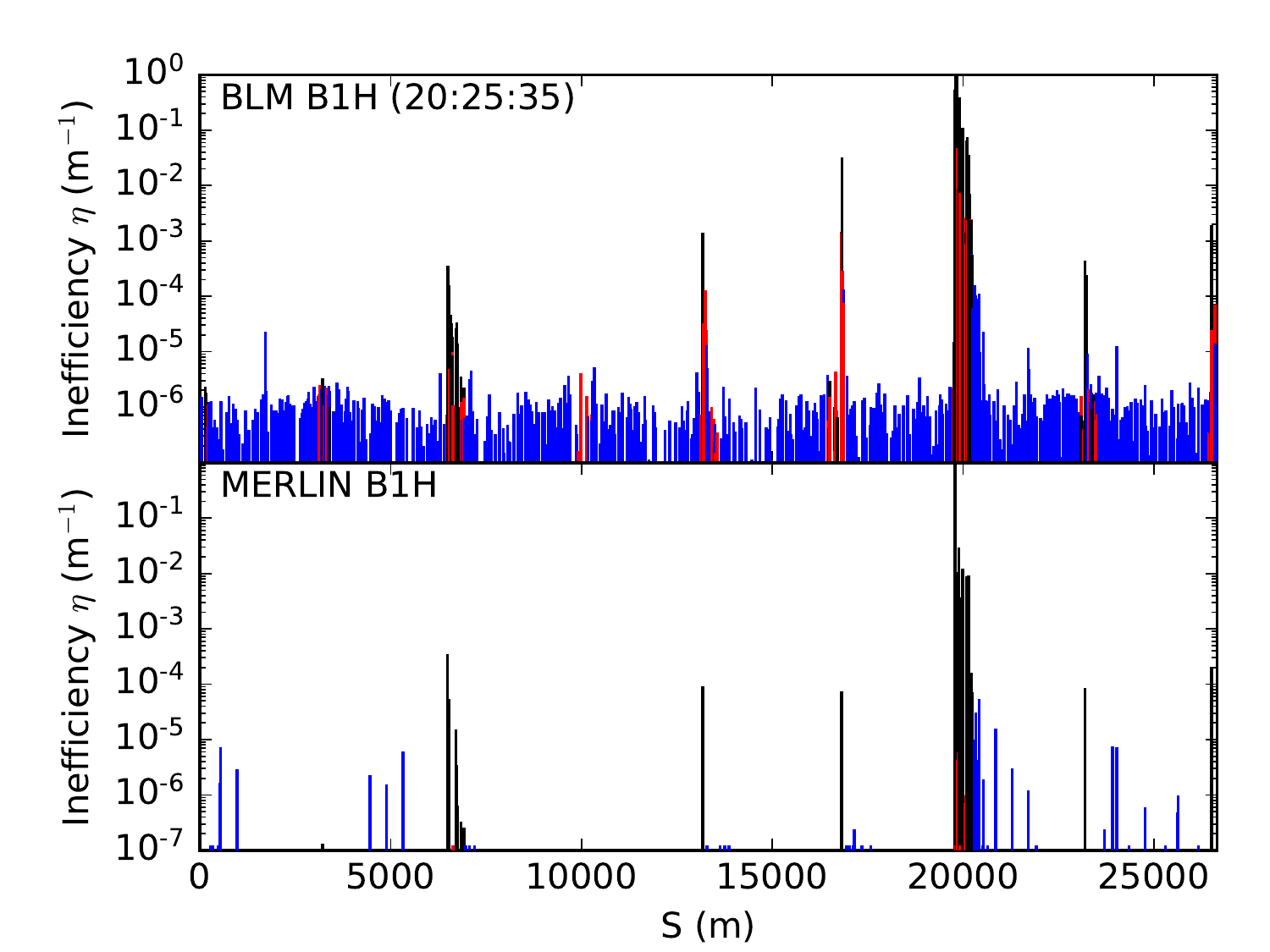}
  \caption{Beam 1 horizontal loss map from BLMs (top) and Merlin++ (bottom) at 6.5 TeV with $\beta$* of 50 cm.~\cite{tygier_performance_2019}}
\label{fig:lossmap_2016_Ring_B1H}
\end{figure}

In \cite{tygier_performance_2019} losses were compared to BLM data under a wide range of LHC operational parameters at both 4~TeV and 6.5~TeV. Figure \ref{fig:lossmap_2016_Ring_B1H} shows a recorded and a Merlin++ simulated loss map. Merlin++ reproduces all the main loss locations as well as the collimation hierarchy, the relative losses at the primary, secondary and tertiary collimators. The article also includes comparisons of loss ratios for a range of intermediate optics points during the beam squeeze phase.

\section{Profiling and Performance}

\label{sec:speed}

\subsection{Fast Computation Design Features}

Merlin++ is designed to be a high performance code, capable of tracking large numbers of particles through lattices with many magnets. Data structures are designed to reduce unneeded storage overhead, and simulation code  avoids unnecessary computation or data copying. Improvements are ongoing and changes are made where profiling shows that they will be effective.
For example, the {\tt RTMap} class for second order transfer maps stores only a sparse matrix of the non-zero elements to reduce unnecessary computations.

\subsection{Multi-threading}

For the most common use cases of Merlin++ only single particle dynamics are considered. This allows many Merlin++ simulations to be split into completely independent tasks, each with a subset of particles. These can be run as separate processes on one or more computer nodes. Merlin++ supports the use of Message Passing Interface (MPI) parallelism for this purpose. The output from each process can then be combined with a post processing script if needed, for example to sum the losses in each element.

\begin{figure}[htb]
  \includegraphics[width=0.46\textwidth]{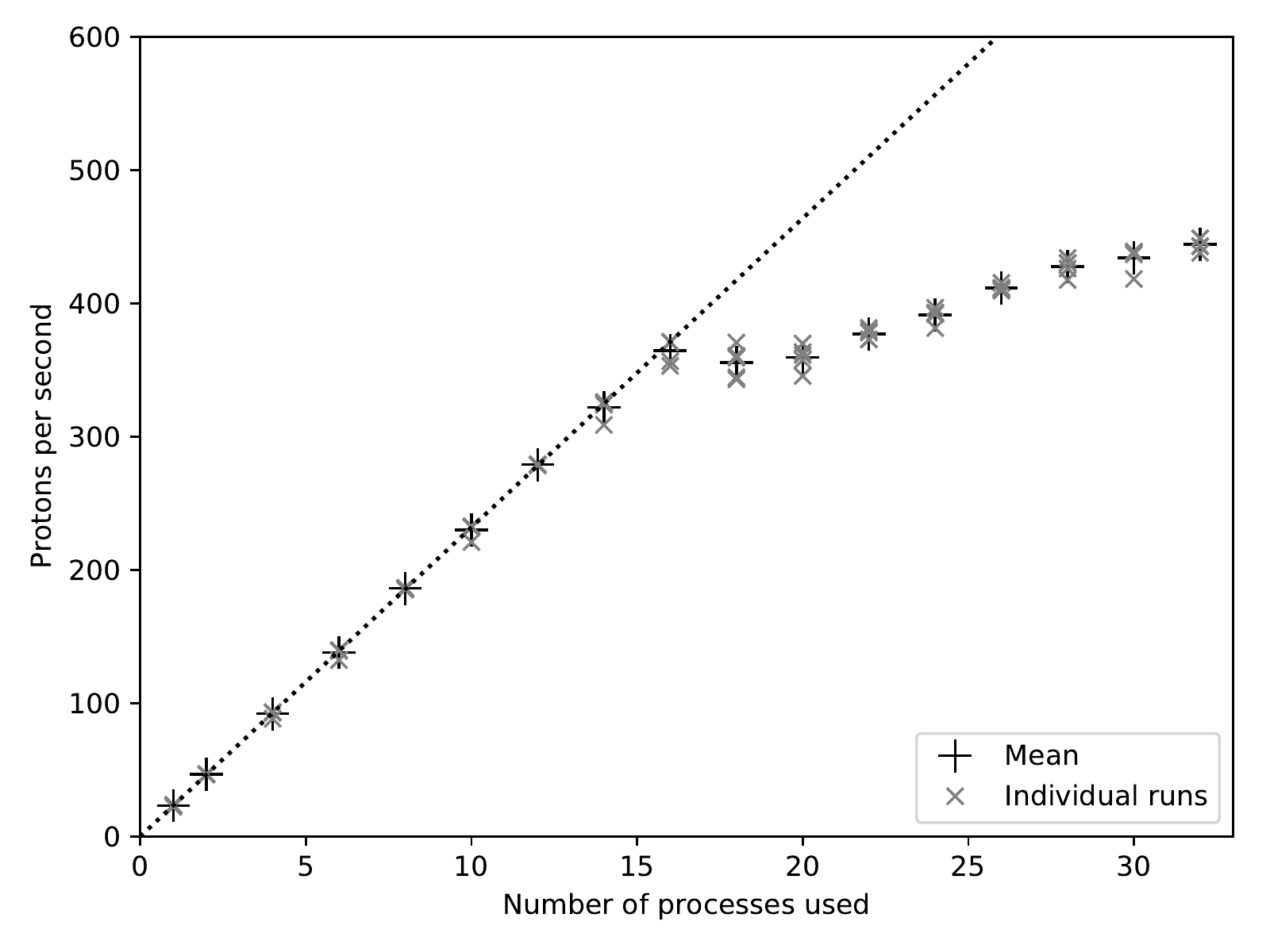}
  \caption{Performance of a benchmark test case with number of processes on a 16-core 32-thread Xeon processor. Dashed line shows ideal scaling.}
\label{fig:multi_proc_benchmark}
\end{figure}

Figure~\ref{fig:multi_proc_benchmark} shows the throughput from running an LHC 200 turn loss map simulation with multiple processes. The benchmark was run on a dual Intel Xeon E5-2650 v2 system with 16 cores and 32 threads using hyper threading, with 10k particles per process. The performance increase is close to the expected ideal scaling up to the 16 real cores, beyond that the gain is smaller due to shared resources in hyperthreading.

This method works well with high throughput batch systems such as HTCondor~\cite{condor-practice} which allows large clusters to be formed even from inhomogeneous and intermittently available compute nodes.

\begin{figure}[htb]
  \includegraphics[width=0.48\textwidth]{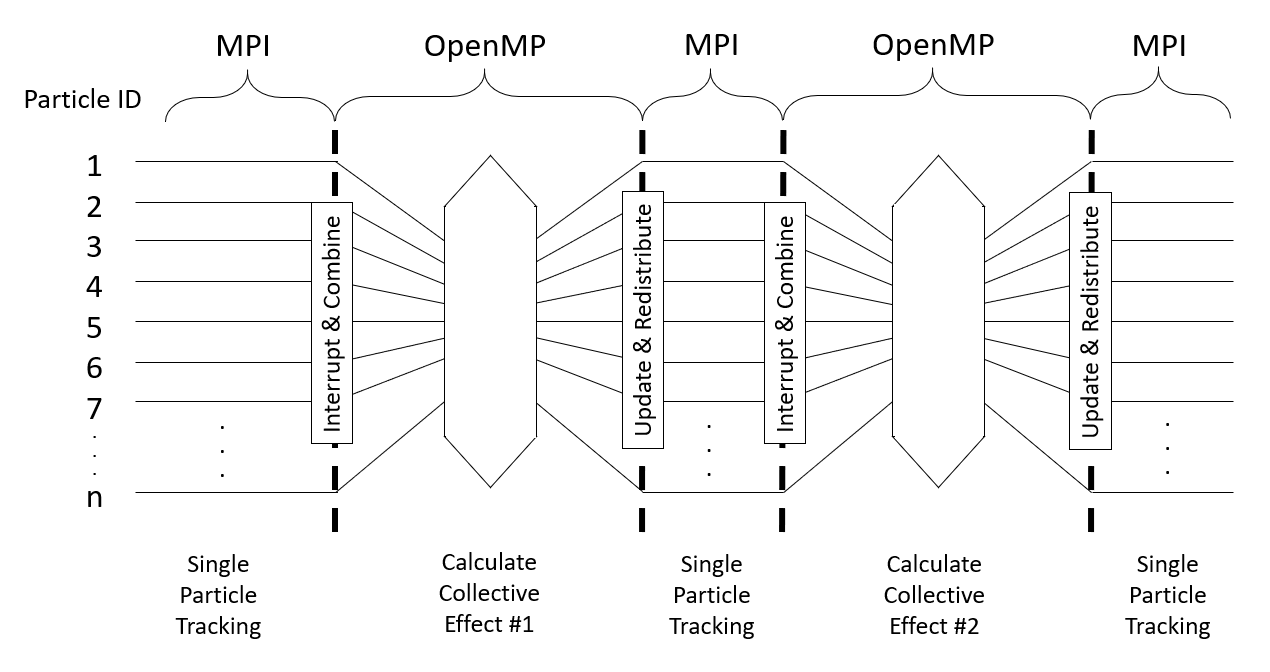}
  \caption{Graphical depiction of how Merlin++ allows the combination of both distributed single particle tracking (MPI) and calculation concurrency (OpenMP) to maximise concurrency potential in taking into account collective  effects.}
\label{fig:collective_parallel}
\end{figure}

For cases where multi-particle or bunch-wide dynamics are required communication between processes must be taken into account, {\em e.g.} tracking sliced macro-particle bunches in wake fields simulations. For this purpose, Merlin++ has built-in support for OpenMP parallelism. Merlin++ allows the user to define interrupts/halts for individual particle trackers at a given process stage. The user may then combine all particle data to carry out a collective bunch-wide calculation utilizing multi-threading where possible without resulting in data races. Merlin++ can then redistribute updated individual particle data to continue the tracking process. This method of parallelism for collective effects is depicted in figure~\ref{fig:collective_parallel}. Performance of Merlin++ running in OpenMP mode depends strongly on the amount and frequency of inter-process communication needed for the simulation and the cluster hardware. For OpenMP use-cases, the developers are investigating the use of C++17 standard library native parallel algorithms for a future release.

\section{Using Merlin++}
\label{sec:using}
\subsection{Installation and Compilation}

Merlin++ is distributed in source code form and is compiled by the user. 

Current system requirements are a C++ compiler (ideally {\tt g++}), CMake and {\tt git}. Python is needed to run the initial tests, which though not strictly essential is strongly advised. 
If Eclipse is being used as an IDE then the Java Runtime Environment is needed. These are all standard and exist on most systems used for scientific calculation. The CERN ROOT package is 
useful but not essential. The user creates and initialises a {\tt git} directory and clones Merlin++ into it from the repository. They then create a build directory and run {\tt CMake} and {\tt make} to compile the library. A test suite is run to verify a successful download, and the user can then start
writing and developing their own programs.

Full (and up to date) installation instructions can be found on the Merlin++ website~\cite{merlin_website}. 

%

\subsection{User Programs}
A Merlin++ user program is written in C++ and linked with the Merlin++ library prior to execution. While the possibilities for a user program are large, a typical program will usually contain the following steps: 
\begin{itemize}
    \item 
Load an accelerator lattice description from a file or build one programmatically.
\item Define or load an initial bunch of particles
\item Set up a tracker with appropriate physics processes
\item Run the tracker
\item Perform analysis on and/or output the results. 
\end{itemize}

\begin{figure} [htb]
  \includegraphics[width=0.5\textwidth]{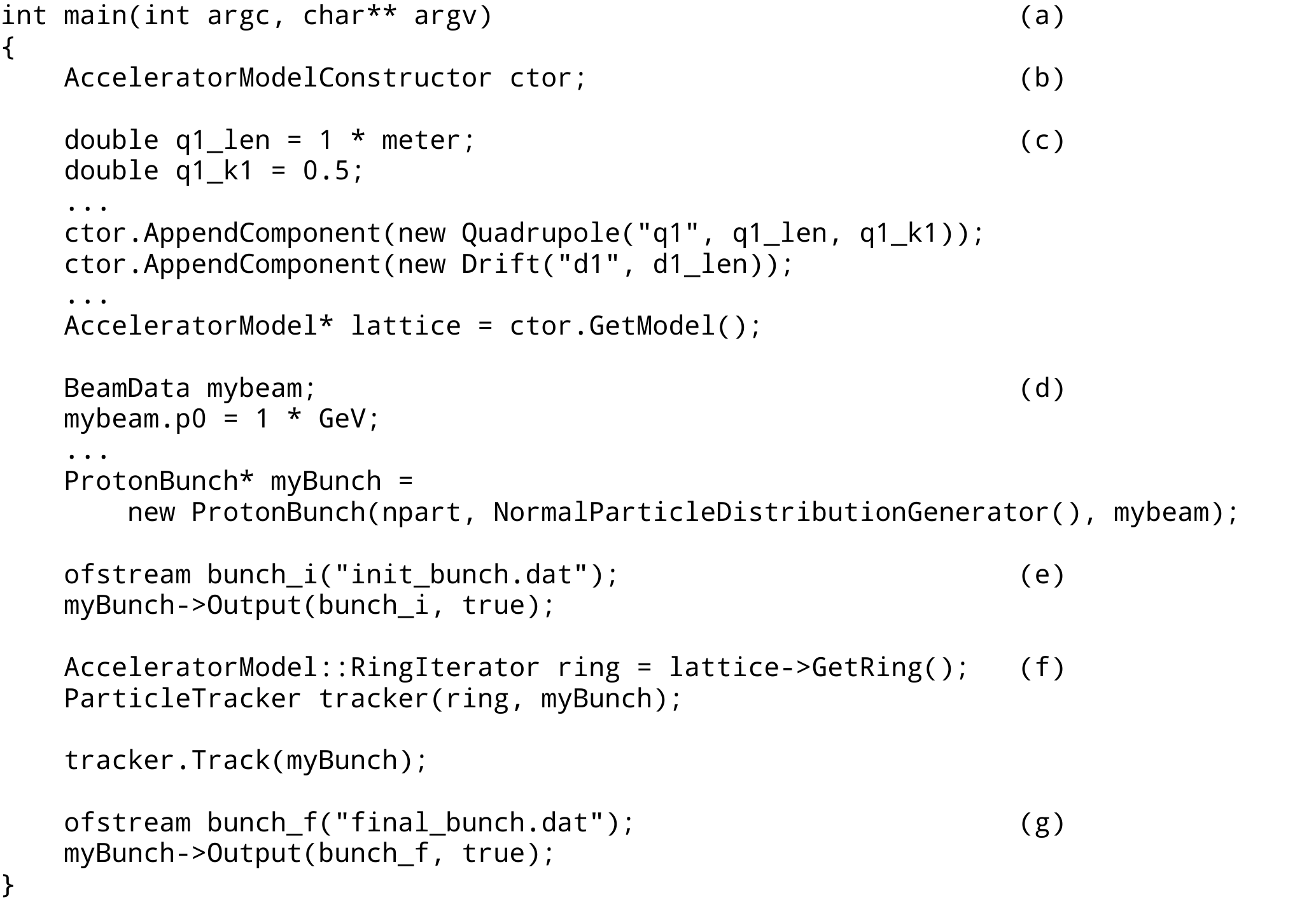}
  \caption{A simplified example of a user program. The program is standard C++ (a). First an accelerator lattice is constructed (b) which can be parametrized (c). Then the beam parameters can be defined and a particle bunch constructed (d). The initial bunch is written to disk (e), then tracked though the lattice (f) and the resulting particle coordinates written to disk (g).}
\label{fig:user_example}

\end{figure}

A code snippet showing the structure of a basic Merlin++ simulation is provided in Figure~\ref{fig:user_example}. More advanced use-cases may include dynamic modification of model elements, interfacing with external programs and/or iteration of beam/optics parameters over multiple simulation runs. A number of examples for common cases are provided with the Merlin++ source code. Furthermore, an accompanying {\em `Quick-Start Guide'} provides a detailed walk-through for installation and compilation and a series of tutorials on user program development, including tracking, lattice construction/import and manipulation and physics process activation. Corresponding Python \cite{python} output and analysis scripts are also provided for each tutorial. 

\subsection{Defining a Lattice}
A lattice structure can either be defined manually by using the native API or automatically via importing a MAD X {\tt .tfs} file, which is done using the {\tt MADInterface} class. 
For manual construction, the user must first define a new model, which is most easily done by instantiating the {\tt AcceleratorModelConstructor} container class and calling its member function {\tt NewModel()}. An accelerator model can then have {\tt AcceleratorComponents} appended via the {\tt AppendComponent()} method which creates and assigns a new instance of a specified component. A component constructor typically requires a reference ID as well as length and field parameters, $k_1$, $k_2$, etc., {\em e.g.} {\tt Quadrupole(`q1',q1\_len,q1\_k1)}. To make the construction of a periodic lattice which contains multiple identical lattice cells in succession one can simply implement a series of {\tt AppendComponent} commands in a loop. To finalise the lattice definition an instance of the {\tt AcceleratorModel} class must be created which accepts an {\tt AcceleratorModelConstructor} as an input, using the {\tt GetModel()} member function. The {\tt AcceleratorModel} class comes defined with functional iterators {\tt ::RingIterator}, to cycle through the lattice components if required for analysis. This iterator is also used by the tracker itself. 

\subsection{Modifying Lattice Elements}
\begin{figure} [htb]
\centering
  \includegraphics[width=0.45\textwidth]{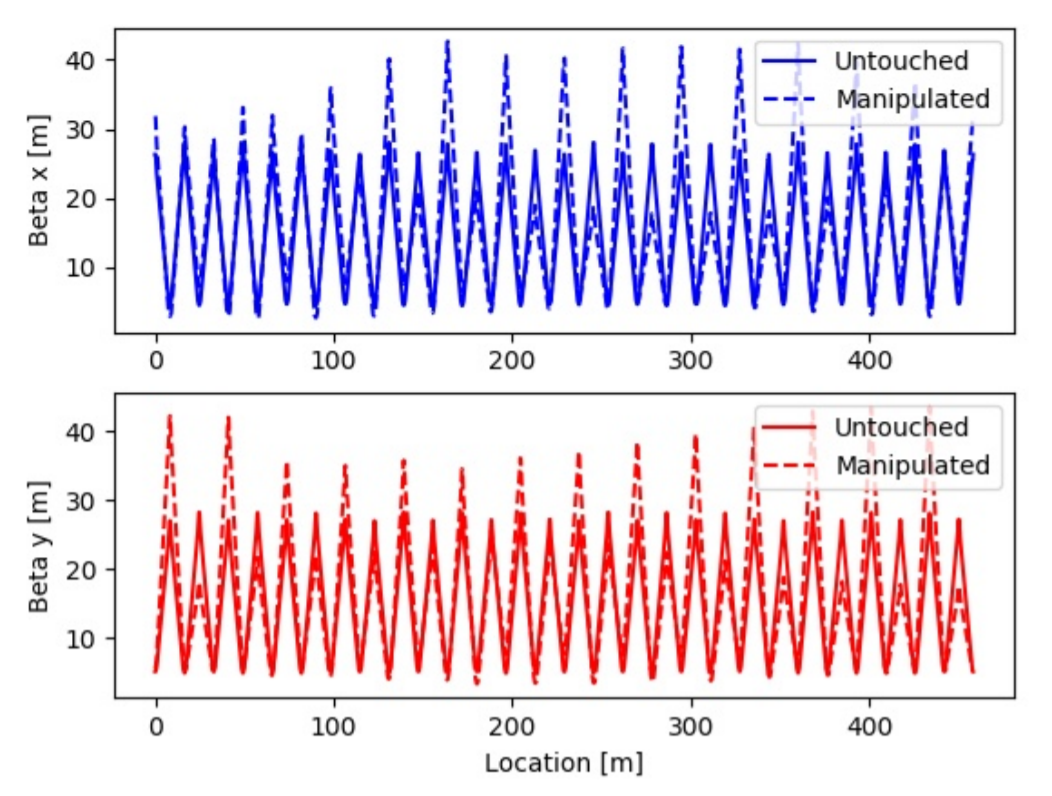}
  \caption{An example of how modifying specific lattice element position and field strengths affect the beta function. Iterative modification can be implemented to achieve automated design of stable lattices.}
\label{fig:lattice_modification}
\end{figure}

Specific lattice elements can be modified without having to redefine the whole lattice. Depending on how the lattice is defined the program may already have pointers to the element of interest or it can be found with searches on name or type, for example with {\tt AcceleratorModel::ExtractTypedElements} class method. To offset a magnet’s alignment, a user would extract a {\tt MagnetMoverList}, then use the {\tt magnet[n]->SetX()} and {\tt magnet[n]->SetY()} member functions to alter the transverse position of magnet $n$. To alter the field one can use {\tt GetFieldStrength()} method which is defined for standard magnets or directly access the field using {\tt AcceleratorComponent::GetEMField()}. An example of lattice modification is shown in figure~\ref{fig:lattice_modification}. Lattice modification can be done iteratively within a loop to achieve fine-tuning in lattice design stability.

\subsection{Defining Element Apertures}
Real-time aperture boundary checks can be used to simulate the presence and limitations of a physical vacuum pipe. Physical boundaries are assigned to specific elements using the {\tt ApertureConfiguration} class. The class constructor accepts a MAD-style {\tt .tfs} file input which specifies apertures using the MAD  parameters APER\_1-4. This allows the definition of multiple aperture shapes, the options being  {\tt RectangularAperture}, {\tt CircularAperture}, {\tt EllipticalAperture}, {\tt RectEllipseAperture}, and {\tt OctagonalAperture}. (If a user requires a shape other than these it would be simple for them to write an appropriate class definition.)  Following successful import of the aperture parameters, the user then calls the {\tt ::ConfigureElementApertures()} method. 

\subsubsection{Collimator Elements}
Collimators can be included as part of a lattice structure. To add a series of collimators at specified locations, The user provides a collimator database in a text file. This file contains collimator component names, locations, lengths and  materials. A {\tt CollimatorDatabase} class constructor accepts this database file as an input argument and collimator information is subsequently added to the existing lattice structure upon calling the {\tt ::ConfigureCollimators()} class method. In each component instance a {\tt CollimatorAperture} class is instantiated with information on the jaw gap (using the aperture class discussed in the previous section) and the material.

\subsubsection{Collimator Materials}
\label{sec:materials}
To specify collimator properties there is a class {\tt MaterialProperties}. 
For all materials basic properties such as density are provided. For more specific use-cases, such as resistive wakefields which require to know the conductivity, parameters are not created by default but can be added. The class constructor for {\tt MaterialProperties} takes 8 arguments: $A$ the atomic weight, $\sigma_T$, $\sigma_I$, and $\sigma_R$, the total, inelastic and Rutherford cross sections,  $dE/dx$ the  energy loss, $X_0$ the radiation length, $\rho$ the  density and $Z$ the atomic number, and these can also be specified by the 
{\tt ::SetMaterialProperties()} method. For properties outside the standard set, the member function {\tt SetExtra()} allows additional parameters to be defined.

For scattering simulations, the cross sections are important. The total and inelastic cross sections, $\sigma_T$ and $\sigma_I$, are specified through {\tt MaterialProperties}. Formul\ae\ do exist which predict them, however they do not always agree with experimental values. As such the user has been given explicit control.

Although some users will want to specify all material properties of a component  themselves, most will just want to specify a common material, such as aluminium, copper, or stainless steel. For this purpose, a native {\tt MaterialData} class is provided. Its most common use is as the child class {\tt StandardMaterials} which fills a map whose keys are meaningful character strings for commonly-used materials such as {\tt "Cu"} and {\tt "Fe"}, and whose values are the appropriate {\tt MaterialProperties}.

Material mixtures are specified by reference to entries in the {\tt MaterialData} map. For example
\begin{verbatim}
StandardMaterialData* matter =
    new StandardMaterialData;
matter->MakeMixture("IT180","W Ni Cu",
   0.95,0.035,0.015,18.06*gram/cc);
\end{verbatim} 
will create a mixture of Tungsten, Nickel and Copper in the numerical ratios 95:3.5:1.5. (The fractions do not have to add to unity as the numbers are subsequently normalised.) The density has to be specified as it cannot be deduced from the constituent properties. For convenience
a {\tt MakeMixtureByWeight} is also provided.

\subsection{Defining the Beam}
A beam can be defined either by specifying individual particle properties or by defining a bunch distribution. In each case an instance of the {\tt ParticleBunch} class is created. For individual particles, the {\tt Particle} class allows the user to specify phase-space information, as defined in \ref{appendixone}. Individual particles are added to the bunch via the {\tt ::AddParticle(Particle)} method. 

To specify particle bunches as a whole, the {\tt ParticleDistributionGenerator} class allows a particle bunch of typical distributions to be constructed. Supported distributions include: uniform, normal/Gaussian, ring, Landau, and Hollow (for HEL simulations). In each case the distribution is determined via random number seeding, outlined in the following section. Also beam parameters must be specified via the {\tt BeamData} class, as it is not inherent. {\tt BeamData} members include: energy {\tt p0}, {\tt charge}, Courant-Snyder parameters {\tt alpha} and {\tt beta}, beam energy and length spreads {\tt sig\_dp} and {\tt sig\_dp}, and transverse emittances {\tt emit\_x} and {\tt emit\_y}. 

\subsubsection{Random Number Seeding}

When using Merlin++ it is often possible to split the simulation into multiple sub-tasks. For example, a collimation loss map simulation might use $10^8$ initial particles, which could be tracked in $100$ batches of $1,000,000$ particles (due to the independence between particles). In this case, it is important that the random number streams used are uncorrelated between each batch. This achieved by giving each batch a different initial random seed. It is recommended that a user explicitly defines the seeding of the random number generator, however, if this is not done then a seed is automatically generated using {\tt std::random\_device} -- considered a high quality random source on most systems. The seed can be set by passing one or more unsigned 32~bit integers to the {\tt RandomNG::init} method. Using multiple values as seed can be useful to avoid statistical correlations between related simulations. For example, when using multiple simulation parameters and multiple batches of particles, then the parameter set number and batch number could both be passed as seed values.

Merlin++ does not aim to guarantee bit for bit reproducibility between multiple runs with the same input. Differences between compilers, compiler options, system libraries and CPU architectures can give small differences to floating point operations. While most modern platforms conform (or can be set to conform) to the IEEE 754 standard for floating-point arithmetic, there are still known discrepancies. Excess precision can occur where intermediate steps of a calculation are not rounded to 64~bit. On modern CPUs, this can happen with fused-multiply-add operations. Minor differences in the implementations of the random number distributions in the C++ standard library are therefore expected.

\subsection{Running a Simulation}
Running a tracking simulation, following configuration of the lattice and beam parameters, takes only a few lines of user code. A {\tt ParticleTracker} class constructor accepts the lattice and particle bunch data as inputs, {\em e.g.} 
\begin{verbatim}
ParticleTracker 
    tracker(myLattice->GetBeamline(), myBeam)
\end{verbatim}
A user can then run the tracker simply calling the member function tracker.Run().

\section{Summary}
\label{sec:conclusions}
Merlin++ continues to be in active development and use. 
The library contains an extensive set of tools for constructing detailed accelerator models, including standard component types, alignment and geometry information, apertures, and collimator material properties.
It contains an accurate high performance tracking integrator, as well as a collection of physics processes needed for advanced accelerator design and optimization. It uses modular object oriented design so that new physics processes can be added as needed by users.

Several features are being considered for future releases.
Ion tracking is needed for simulation of accelerators such as RHIC.
Variable discretised Truncated Power Series Algebra will enable fast accurate symplectic long-term lifetime studies.
An implicit symplectic Runga Kutta integrator would give very accurate tracking, which might be slow but could be used as a benchmark for TPSA accuracy.
It will continue to adopt new features in emerging programming standards (such as C++17) 
to ensure high performance and high quality coding standards.
As accelerator hardware become technically more sophisticated to meet future challenges of energy, luminosity, and precision, Merlin++ will continue to provide the simulation facilities needed to ensure successful design and operation of new machines. 

\section{Acknowledgements}

Many people have contributed to the  Merlin++ source code and we would like to acknowledge significant contributions from : Adriana Bungau, James Fallon,  Maurizio Serluca, Kyrre Ness Sjøbæk and  Adina Toader. The current phase of development has been funded by
CERN and STFC as part of the HL-LHC project.

\vfill\eject
\appendix
\section{The definition of particle parameters}

\label{appendixone}

Each particle tracked is specified by 
7 numbers, the contents of {\tt PSVector} and also known, through a {\tt typedef}, as {\tt Particle}

\begin{enumerate}
    \item {\tt x}, the horizontal transverse co-ordinate in metres
    \item {\tt xp}, the canonical $x$ momentum, approximately equivalent to the horizontal transverse angle in radians.
    \item {\tt y}, the vertical transverse co ordinate in metres
    \item {\tt yp}, the canonical $y$ momentum, approximately equivalent to the  vertical transverse angle in radians.
    
    \item {\tt ct}, the longitudinal distance from the bunch centre, in metres
    
    \item {\tt dp}, the fractional deviation of the momentum from the nominal value {\tt p0}
    
    \item {\tt id}, which can be used as a unique particle identifier.
    
\end{enumerate}

   On modern x86-64 CPUs we do not find any performance benefit to making the particle storage size a power of 2 for example by adding an additional coordinate or padding.
    
    Merlin does not strictly enforce these units - the user is free to provide processes using different units provided they are consistent within their simulation.

\section{Transfer maps for the Symplectic Integrator}

\label{appendixtwo}

The accelerator Hamiltonian, $H$, as given in Equation 2.73 of \cite{Wolski1} 
describes the motion of a 
charged particle in a general electromagnetic field, is written with its six degrees of freedom (transverse positions $x$ and $y$ and momenta $p_x$ and $p_y$, energy $E$ and longitudinal distance travelled $s$) converted to canonical co-ordinates using a fixed reference momentum, $P_{0}$.
This is used to derive the transfer maps.

For a drift

\begin{equation}
x(s)=x+\frac{p_{x} s}{d},
\end{equation}
\begin{equation}
p_{x}(s)=p_{x},
\end{equation}
\begin{equation}
y(s)=y+\frac{p_{y} s}{d},
\end{equation}
\begin{equation}
p_{y}(s)=p_{y},
\end{equation}
\begin{equation}
z(s)=z+\frac{s}{\beta_{0}}\Big(1-\frac{1}{d}\Big)-\frac{\delta s}{d},
\end{equation}
\begin{equation}
\delta(s)=\delta,
\end{equation}
\noindent where
\begin{equation}
d = \sqrt{\Big(\delta+\frac{1}{\beta_{0}}\Big)^2 - p_{x}^2 - p_{y}^2 - \frac{1}{\beta_{0}^2 \gamma_{0}^2}}.
\end{equation}
and $\beta_0$ and $\gamma_0$ are the relativistic factors for the reference momentum.
\noindent Note that due to the square root component, $d$, the resultant exact drift transfer map is inherently non-linear (this is not the case for {\tt THIN\_LENS} and {\tt TRANSPORT} integrators). 

The implementation of the Forest sector bend transfer map follows:
\begin{equation}
x(s) = \frac{1}{b_{1}}\Big(\sqrt{(1+\delta)^2-p_{x}(s)^2-p_{y}^2}-\rho \big( \frac{d p_{x}(s)}{d s} - b_{1}\big)\Big)
\end{equation}
\begin{equation}
p_{x}(s) = p_{x} \cos\left(\frac{s}{\rho}\right)+\left(\sqrt{(1+\delta)^2-p_{x}^2-p_{y}^2}-b_{1}(\rho + x)\right) \sin{\left(\frac{s}{\rho}\right)}
\end{equation}
\begin{multline}
y(s) = y+\frac{p_{y} s}{b_{1} \rho}+ \\ \frac{p_{y}}{b_{1}}\left(\sin^{-1}\left(\frac{p_{x}}{\sqrt{(1+\delta)^2-p_{y}^2}}\right)-\sin^{-1}\left(\frac{p_{x}(s)}{\sqrt{(1+\delta)^2-p_{y}^2}}\right)\right)
\end{multline}
\begin{equation}
p_{y}(s)=p_{y}
\end{equation}
\begin{equation}
\delta(s)=\delta
\end{equation}
\begin{multline}
z(s) = z+\frac{(1+\delta) s}{b_{1} \rho}+\\
\frac{(1+\delta)}{b_{1}}\left(\sin^{-1}\left(\frac{p_{x}}{\sqrt{(1+\delta)^2-p_{y}^2}}\right)-\sin^{-1}\left(\frac{p_{x}(s)}{\sqrt{(1+\delta)^2-p_{y}^2}}\right)\right).
\end{multline}
\noindent In this case, $b_1$ is a dipole field constant and $\rho$ is the curvature of dipole. The rectangular dipole equivalent, assuming ideal parallel faces, is then derived by taking the limit $\rho \xrightarrow \, \infty$.

\bibliography{merlin.bib}{}
\bibliographystyle{elsarticle-num}

\end{document}